\documentclass[11pt]{article}
\usepackage{jcapmod}

\usepackage{booktabs}
\usepackage[english]{babel}
\usepackage{amsmath, amssymb, amsbsy, amstext, amsthm, simplewick}
\usepackage{hyperref}
\usepackage{graphicx}
\usepackage{amsfonts}
\usepackage{upgreek}
\usepackage{framed}
\usepackage{pifont}
\usepackage{latexsym, mathrsfs}
\usepackage{array}
\usepackage{hyperref}
\usepackage{xspace}
\usepackage{longtable}
\usepackage{multirow}
\usepackage{cases}
\usepackage{empheq}
\usepackage[lofdepth,lotdepth]{subfig}

\usepackage{braket}

\usepackage[load-configurations=astronomy, range-units=brackets, range-phrase=-, per-mode=reciprocal, mode=math]{siunitx}
\usepackage{threeparttable}
\usepackage{subfig}

\definecolor{Blue}{rgb}{0.25, 0.41, 0.88}
\definecolor{Red}{rgb}{0.92,0.,0.}
\definecolor{darkorange}{rgb}{1.0,0.549,0.}
\definecolor{pyorange}{RGB}{255,140,0}
\definecolor{pybrown}{RGB}{149,100,0}
\definecolor{pydarkred}{RGB}{209,0,0}
\definecolor{pymidblue}{RGB}{0,128,255}
\definecolor{pyyellow}{RGB}{255,213,0}
\definecolor{regionred}{RGB}{255,228,225}

\newcolumntype{C}[1]{>{\centering\let\newline\\\arraybackslash\hspace{0pt}}m{#1}}

\def\d{{\rm d}}
\def\l{{\ell}}

\def\r{{\bf r}}
\def\R{{\bf R}}

\def\p{{\bf p}}

\makeatletter
\newlength{\apb@width}
\newcommand{\autoparbox}[2][c]{\settowidth{\apb@width}{#2}\parbox[#1]{\apb@width}{#2}}

\makeatother

\numberwithin{equation}{section}

\def\beq{\begin{equation}}
\def\eeq{\end{equation}}

\def\bea{\begin{eqnarray}}
\def\eea{\end{eqnarray}}

\def\d{{\rm d}}

\def\beq{\begin{equation}}
\def\eeq{\end{equation}}
\def\bea{\begin{eqnarray}}
\def\eea{\end{eqnarray}}

\def\d{{\rm d}}

\def\d{{\rm d}}
\def\l{{\ell}}

\def\R{{\bf{R}}}
\def\r{{\bf{r}}}

\DeclareRobustCommand{\SkipTocEntry}[4]{}

\RequirePackage{color}
\usepackage{colortbl}
\definecolor{blue2}{cmyk}{1, 0.1, 0.1, 0.1}

\usepackage{hhline}
\usepackage{array}
\newcolumntype{P}[1]{>{\centering\arraybackslash}p{#1}}
\usepackage[usenames,dvipsnames,svgnames,table]{xcolor} 
\usepackage{booktabs}
\usepackage{tikz}
\usetikzlibrary{calc,shadings,patterns,tikzmark}

\begin{document}

\pagenumbering{roman}
\begin{titlepage}
\baselineskip=15.5pt \thispagestyle{empty}

\begin{flushright}
DESY 18-060
\end{flushright}
\bigskip\

\bigskip\

\vspace{1cm}
\begin{center}
{\fontsize{18}{24}\selectfont  \bfseries  {
Probing Ultralight Bosons with Binary Black Holes}} 
\end{center}
\vspace{0.1cm}
\begin{center}
{\fontsize{12}{18}\selectfont Daniel Baumann,$^{1}$ Horng Sheng Chia,$^{1}$ and Rafael A.~Porto$^{2,3,4}$} 
\end{center}

\begin{center}
\vskip8pt
\textsl{$^1$ Institute of Theoretical Physics, University of Amsterdam,\\Science Park 904, Amsterdam, 1098 XH, The Netherlands}

\vskip8pt
\textsl{$^2$   ICTP South American Institute for Fundamental Research, \\ 
Rua Dr. Bento Teobaldo Ferraz 271, 01140-070 Sao Paulo, SP Brazil}\vskip 8pt
\textsl{$^3$  Deutsches Elektronen-Synchrotron DESY,\\ Theory Group, D-22603 Hamburg, Germany}\vskip 8pt
\textsl{$^4$ Max Planck Institut f\"ur Gravitationsphysik (Albert Einstein Institute),\\ Callinstr. 38, D-30167 Hannover, Germany}
\end{center}

\vspace{1.2cm}
\hrule \vspace{0.3cm}
\noindent {\bf Abstract}\\[0.1cm]
We study the gravitational-wave (GW) signatures of clouds of ultralight bosons around black holes (BHs) in binary inspirals.  These clouds, which are formed via superradiance instabilities for rapidly rotating BHs, produce distinct effects in the population of BH masses and spins, and a continuous monochromatic GW signal. We show that the presence of a binary companion greatly enriches the dynamical evolution of the system, most remarkably through the existence of resonant transitions between the growing and decaying modes of the cloud (analogous to Rabi oscillations in atomic physics). These resonances have rich phenomenological implications for current and future GW detectors. Notably, the amplitude of the GW signal from the clouds may be reduced, and in many cases terminated, much before the binary merger. The presence of a boson cloud can also be revealed in the GW signal from the binary through the imprint of finite-size effects, such as spin-induced multipole moments and tidal Love numbers. The time dependence of the cloud's energy density during the resonance leads to a sharp feature, or at least attenuation, in the contribution from the finite-size terms to the waveforms. The observation of these effects would constrain the properties of putative ultralight bosons through precision GW~data, offering new probes of physics beyond the Standard~Model. 
\vskip10pt
\hrule
\vskip10pt

\end{titlepage}

\thispagestyle{empty}
\setcounter{page}{2}
\tableofcontents

\clearpage
\pagenumbering{arabic}
\setcounter{page}{1}

\clearpage
\section{Introduction}
\label{sec:introduction}

The recent detections of gravitational waves (GWs) by the LIGO/Virgo collaboration~\cite{Abbott:2016blz, Abbott:2016nmj, Abbott:2017vtc, Abbott:2017oio,TheLIGOScientific:2017qsa} mark the beginning of multi-messenger astronomy \cite{GBM:2017lvd} and the birth of `precision gravity' \cite{Porto:2016zng,Porto:2017lrn}. Binary systems, of comparable masses or extreme-mass ratios, will become the leading probe to test gravitational dynamics and the physics of compact objects, such as black holes (BHs) and neutron stars (NSs), under unique conditions. While it is indisputable that future GW observations will play a transformative role in~astrophysics \cite{Buonanno:2014aza}, it is less clear what impact these measurements will have on other branches of physics, and in particular, whether they can shed light on phenomena beyond the Standard Model. In principle, the non-perturbative regime of gravitational dynamics may carry imprints of `new physics', see e.g.~\cite{Palenzuela:2017kcg}. However, the need for numerical modelling, together with the small number of cycles involved, may hinder our ability to pinpoint different scenarios, if restricted only to the merger phase. In contrast, an accurate analytic reconstruction of the signal during the inspiral, in combination with simulations for the late stages of the coalescence, offers a unique opportunity to study physics beyond the Standard Model through GW precision~data.

\vskip 4pt 
Using GW observations as probes of new physics is challenging, mainly due to the efficient decoupling of short-distance physics from long-distance~observations. 
Indeed, finite-size effects, characterized by higher-derivative terms in a `worldline' effective field theory (EFT) approach~\cite{nrgr,nrgrs,dis1,dis2,Goldberger:2007hy,Porto:2007pw,andirad,andi,Foffa:2013qca,eftgrg20,Rothstein:2014sra,Porto:2016pyg}, encapsulate the physics at scales shorter than the size of the objects sourcing GWs in a binary system. Because Einstein gravity is derivatively coupled, these terms scale with high powers of the ratio between the size and the separation of the bodies. For a non-rotating compact object, the first correction to the structureless point-particle approximation is due to tidal effects and scales with the fifth power of its size. It is often parametrized in terms of the so-called (tidal)  `Love numbers', e.g.~\cite{Binnington:2009bb,Damour:2009vw}, which are analogous to susceptibilities in electrodynamics. These Love numbers modify the phase of the GW signal for inspiraling binary systems at fifth Post-Newtonian (5PN) order~\cite{Flanagan:2007ix, Hinderer:2007mb}. For rapidly rotating bodies, finite-size effects become relevant already at 2PN, through intrinsic spin-induced multipole moments~\cite{Poisson:1997ha,nrgr,nrgrs2}. In either case, the challenge is to extract the parameters of the source accurately \cite{Krishnendu:2017shb, Cardoso:2017cfl, Sennett:2017etc}, which requires very precise waveforms.\footnote{The current state-of-the-art in analytic computations is approaching 4PN order~\cite{Blanchet:2006zz, prl,nrgrss,nrgrs2,nrgrso,srad1,srad2, Levi:2016ofk, Foffa:2012rn,Damour:2014jta,Galley:2015kus,Bernard:2015njp,Damour:2016abl,Foffa:2016rgu,Damour:2017ced,Porto:2017dgs,Porto:2017shd,Marchand:2017pir}, moving foward toward the key 5PN threshold \cite{Porto:2016zng,Porto:2017lrn}.}  As we shall see, both the Love numbers and the spin-dependent multipoles may carry the imprint of new degrees of freedom.  The Love numbers in particular offer a unique diagnostic, since they vanish for BHs in Einstein gravity~\cite{Binnington:2009bb,Damour:2009vw,Kol:2011vg,Gurlebeck}, and therefore any non-zero value would point to physics beyond the Standard Model~\cite{Porto:2016pyg,Porto:2016zng,Rothstein:2014sra}. 
 
 \vskip 4pt
New physics may also appear at distances larger than the size of the objects~in~a~binary. This may entail long-distance modifications of General Relativity or extra~fields as in~scalar-tensor theories. In either case, if the scale of new physics is shorter than the separation between the bodies, decoupling still applies, as far as the computation of GW observables is concerned. For example, modifications of General Relativity can be evidenced  during the inspiral phase through higher-derivative corrections to the Einstein-Hilbert action, see~e.g.~\cite{Endlich:2017tqa}. On the other hand, additional degrees of freedom can have effects at longer distances if their Compton wavelength is larger than the size of the compact objects. This will lead, as we shall see, to larger natural values for the finite-size coefficients in the effective theory, enhanced by powers of the ratio between the Compton scale and gravitational radius. In this manner, new physics may manifest itself at scales larger than anticipated from the Standard Model plus Einstein~gravity~alone.

\begin{figure}[t]
\centering
\includegraphics[scale=0.2, trim = 0 0 0 0]{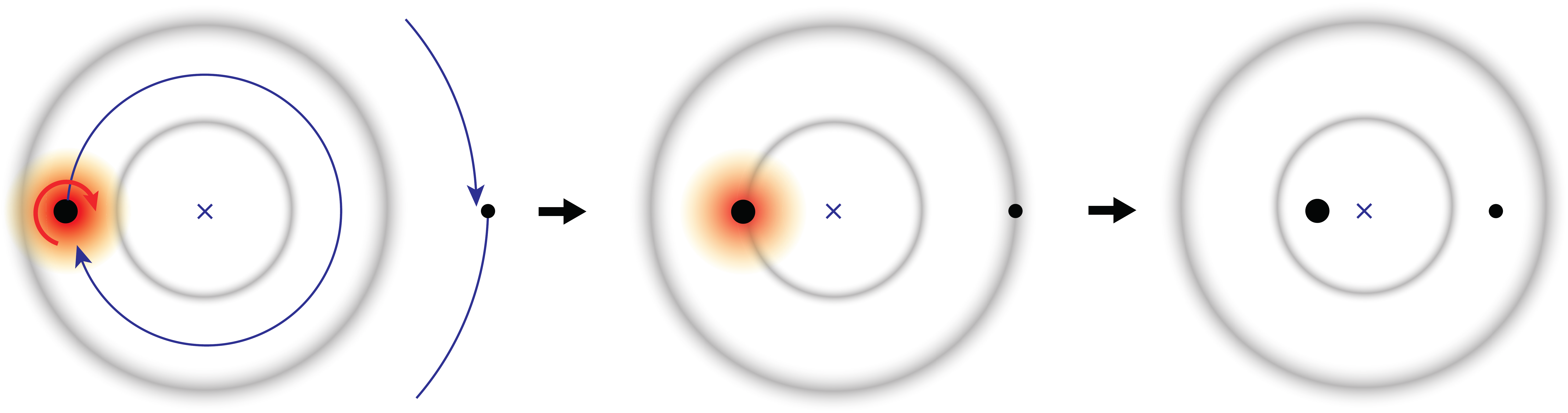}
\caption{Schematic illustration of the evolution of the boson cloud during the inspiral. The system experiences a resonance when the orbital frequency matches the energy difference between growing and decaying modes.  
The orbits for which the resonance occurs are indicated by the gray~bands, marking the point when the cloud may rapidly deplete
leading to novel GW signatures.} 
\label{fig:sketch}
\end{figure}

\vskip 4pt
A notable example for physics beyond the Standard Model are `ultralight' particles~\cite{Essig:2013lka}.
Even though the masses for these particles can be many orders of magnitude smaller than those of known elementary particles, they are technically natural if the coupling to ordinary matter is very weak. At the same time, the required weak coupling makes detecting these particles by traditional experimental means extremely challenging. On the other hand, their large Compton wavelengths means that, if present in nature, they will be efficiently produced by the superradiance instability of rapidly rotating BHs~\cite{1971JETPL..14..180Z, 1972JETP...35.1085Z, 1972BAPS...17..472M, Starobinsky:1973aij, 1974JETP...38....1S,East:2017ovw, PhysRevD.96.024004}. 
 For bosonic fields, superradiance creates a classical condensate which can carry a significant amount of the total mass and angular momentum of the system, see \cite{Brito:2015oca} for a review.  Rather quickly, on cosmological/astrophysical scales, this leads to the BH carrying an `atmosphere', with a large `cloud' of the field co-rotating with the BH (see Fig.~\ref{fig:sketch}).\footnote{These clouds are a form of `BH hair'. We emphasize, however, that this is not a violation of no-hair theorems for BHs, since these configurations are unstable, eventually returning to the Kerr solution. Nevertheless, if the lifetime is long compared to astrophysical or cosmological timescales, then the BH will carry hair for all practical~purposes.} In the non-relativistic limit, the eigenfunctions of the system are determined by a Schr\"odinger-like equation and the whole set up is sometimes referred to as a `gravitational atom'.
 
 \vskip 4pt
The bosons which can be produced in this manner span a vast range of masses, from $10^{-20}$ to  $10^{-10} \,{\rm eV}$, corresponding to BHs of a few to billions of solar masses.  At the upper end of this range we may find particles which play the role of the QCD axion~\cite{PhysRevLett.38.1440, PhysRevLett.40.223, PhysRevLett.40.279}, while those at the lower end offer alternative dark matter candidates, both with real~\cite{Hui:2016ltb} and complex \cite{Li:2013nal,Dev:2016hxv} scalars. In~addition, a plethora of ultralight axion-like particles arise also as a natural consequence of most string compactifications~\cite{Svrcek:2006yi, Baumann:2014nda}. The  `axiverse'~\cite{Arvanitaki:2009fg}, as it is commonly known, includes fields with a wide range of masses, increasing the chances some of them may have formed boson clouds through superradiance instabilities. 

\vskip 4pt  
In this paper, we investigate the GW signatures from boson clouds surrounding BHs, when the latter are part of a binary. As we will show, the existence of a companion greatly enriches the dynamical evolution of the entire system. One of the most important consequences of the gravitational perturbation occurs when the orbital frequency of the binary matches the energy difference between growing and decaying modes of the cloud, leading to resonant transitions between the two states. This effect is analogous to Rabi transitions in the hydrogen atom under the influence of an oscillating external field. The resonances can deplete (or significantly attenuate) the energy density stored in the cloud, leading to rich phenomenological implications and powerful new probes of physics beyond the Standard~Model. 

\vskip 4pt
For real bosonic fields, there are two main sources of GW radiation: {\it i}\hskip 1pt) a continuous monochromatic signal from the cloud~\cite{Arvanitaki:2010sy} and {\it ii}\hskip 1pt)  emission from the two-body system, that includes the BH-cloud system and its companion.  The former depends on the specific configuration for complex scalar fields around BHs, and may be absent for axisymmetric condensates. Both signals, when present, will be affected by the dynamical evolution of the cloud in the binary, in particular as the system evolves through the resonances. Notably, the amplitude of the continuous signal emitted from the cloud may be reduced, and in many cases terminated, much before the binary merger. The boson clouds can also reveal themselves through finite-size effects in the GW signal from the binary during the inspiral, such as spin-induced multipole moment(s) and tidal Love number(s).  
The time-dependence of the cloud during the resonance leads to a sharp feature, or at least an attenuation, in the contribution from finite-size effects to the waveforms. The observation of these phenomenological effects would help us elucidate the nature of the BH-cloud system, and the putative ultralight bosons, through precision GW~data. 

\vskip 4pt 
Using high-precision GW observations to constrain the particle content in the universe resembles the discipline of collider physics. In our case, a scattering process takes place between the rotating BH and an incident wave of the bosonic field.  
The superradiant amplification of the scattered wave is a type of `resonance' that occurs when the Compton wavelength of the field exceeds the size of the BH horizon.  This resonance creates a large number of ultralight bosons in the intermediate state, which, in the case of real scalars, then decay via self-annihilation into 
GWs. These GWs can be thought of as many soft quanta in the final state, whose frequency is given by (twice) the mass of the boson. 
In a binary system, the initial state features an extra body, producing a final state that includes the GWs emitted during the inspiral.  The existence of a `three-body' initial state leads to an additional resonant behavior, opening new decaying channels for the bosonic field. This new configuration enables us to extract more information about 
 the intermediate state. In particular, the coefficients of finite-size terms in the worldline theory carry the imprints of new particles, which, as we will show, can be constrained by GW precision~data. This is analogous to precision tests in particle physics, where constraints on the coefficients of higher-dimension operators probe new physics beyond the Standard Model~\cite{Skiba:2010xn}.
   
 \vskip 4pt
A similar analogy, between collider physics and the imprints of massive particles in cosmological correlation functions, was drawn in \cite{Arkani-Hamed:2015bza}.\footnote{In the case of the `cosmological collider', particles are produced as `resonances', when their Compton wavelength exceeds the cosmological horizon during inflation.  The properties of the new particles in the intermediate state are reflected in the momentum scaling and angular behavior in the soft limits of correlation functions~\cite{Chen:2009zp, Baumann:2011nk, Assassi:2012zq, Chen:2012ge, Noumi:2012vr, Assassi:2013gxa, Flauger:2013hra, Arkani-Hamed:2015bza, Lee:2016vti, Baumann:2017jvh, Kumar:2017ecc, An:2017hlx}.} Taking inspiration from cosmology, we will refer to this nascent discipline in GW science as `gravitational collider physics', which complements other searches for new light particles, in the lab~\cite{Essig:2013lka}, astrophysics~\cite{Raffelt:1996wa} and cosmology~\cite{Abazajian:2016yjj, Baumann:2016wac, Brust:2013xpv}.

\paragraph{Outline} The outline of the paper is as follows: In Section~\ref{sec: Boson Clouds around Kerr Black Holes}, we review how superradiance generates boson clouds around rotating BHs, and compute their energy spectrum. 
In Section~\ref{sec:perturbation}, we study the dynamics of the BH-cloud when it is part of a binary. We show that the gravitational perturbation produced by a companion induces resonant mixing between growing and decaying modes of the cloud. 
 We determine the efficiency of this level mixing to deplete the energy density in the cloud, as a function of the parameters of the system. In Section~\ref{sec:pheno}, we discuss how the novel features in the dynamics of the BH-cloud affect GW signals, either directly from the cloud or the two-body system including the companion. We estimate the relevant finite-size effects, as well as their time dependence induced by the evolution of the cloud. We then briefly discuss the novel type of constraints on ultralight bosons in binary inspirals, through GW searches both from the cloud and the binary system.  
 We conclude, in Section~\ref{sec:Conclusions}, with a summary of our results and a discussion of open problems. We also provide an outlook on future applications of the gravitational collider. A number of appendices contain additional material: In~Appendix~\ref{sec: BH-Cloud System}, we derive various properties of the BH-cloud, including the higher-order (hyper)fine structure of its energy spectrum and  its quadrupole moment. 
In~Appendix~\ref{sec: Gravitational Perturbation Appendix}, we provide an alternative perspective on the absence of a dipole induced by the companion object. Finally, in Appendix~\ref{sec:eft}, we review aspects of the EFT approach for extended objects and introduce the relevant finite-size corrections.

\paragraph{Notation and conventions} Our metric convention is $(-, +, +, +)$. 
We will use natural units, $\hbar = c = G=1$, throughout.  The gravitational radius of an object of mass $M$ is $r_g \equiv GM / c^2 = M$. 
 Properties of the boson cloud will be denoted by the subscript~$c$, e.g.~$M_c$ and $r_c$ are the mass and typical radius of the cloud. We use the subscript $*$ for quantities related to the companion, e.g.~$M_*$ is its mass and $V_*$ is the induced gravitational potential.  Ignoring the small energy loss in GW emission, the mass of the BH-cloud is equal to the initial BH mass, and we use $M$ for both of them. We denote by $J$ the initial BH angular momentum, which is also equal to that of the BH-cloud. We define the BH spin parameters $\tilde a = a/M = J/M^2$.  The mass and spin of the BH at the moment when superradiance saturates are denoted by the subscript $s$, i.e.~$M_s$ and $a_s$. We introduce the dimensionless variables $\kappa$ and $\Lambda$ to describe the spin-induced quadrupole and tidal Love number. For a Kerr BH, we have $\kappa=1$ and $\Lambda=0$.

\newpage
\section{Boson Clouds around Black Holes} 
\label{sec: Boson Clouds around Kerr Black Holes}

When a BH rotates faster than the angular phase velocity of an incoming wave, it amplifies the energy and angular momentum of the field in its vicinity.\footnote{This is the rotational analogue of Cherenkov radiation~\cite{PhysRevD.58.064014}, where spontaneous emission of light occurs when a test particle moves with a speed that is faster than the phase velocity of the medium.}  This superradiance effect~\cite{1971JETPL..14..180Z, 1972JETP...35.1085Z, 1972BAPS...17..472M, Starobinsky:1973aij, 1974JETP...38....1S}  is a natural mechanism to create clouds of ultralight bosons around Kerr BHs (see~\cite{Brito:2015oca} for a review). 
In this section, we will review this phenomenon and the properties of the boson clouds that it generates.

\subsection{Black Hole Superradiance} 

Consider a rotating BH of mass $M$ and spin $J \equiv a M$. A bosonic field with mass $\mu$ and angular frequency $\omega$ experiences a superradiant instability if
\beq
\frac{\omega}{m} <  \Omega_{\rm H} = \frac{a}{2 M r_+}\, , \label{eqn: superradiance condition}
\eeq
where $m$ is the azimutal angular momentum of the field,  $\Omega_{\rm H}$ is the angular velocity of the black hole and $r_+ \equiv M + \sqrt{M^2 - a^2}$ is the size of the event horizon of the black hole (see Appendix~\ref{sec: BH-Cloud System}). 
Superradiant growth requires $m>0$, i.e.~modes that co-rotate with the Kerr black hole.  

In principle, superradiance occurs for both massive and massless fields. However, the mass of the bosonic field plays the crucial role of a reflecting barrier, so that the superradiantly amplified field is reflected back onto the black hole and continuously extracts angular momentum from it. This provides a natural realization of the `black-hole bomb' scenario proposed by Teukolsky and Press~\cite{PRESS:1972aa}. 
The amplitude of the field increases while the black hole spin decreases, until the inequality (\ref{eqn: superradiance condition}) is saturated. The black hole spin at saturation is 
\begin{align}
\frac{a_s}{M_s} = \frac{4 m (M_s \omega_s)}{m^2 + 4(M_s \omega_s)^2} \, , \label{eqn: BH spin saturation}
\end{align}
where $M_s$ and $\omega_s$ are the relevant quantities evaluated after superradiance has ended. 
The existence of black holes with spins above this critical value would rule out ultralight scalar fields in the corresponding mass range~\cite{Arvanitaki:2010sy}.

\vskip 4pt
Although superradiance is a phenomenon that occurs for bosons of any spin~\cite{Starobinsky:1973aij, 1974JETP...38....1S, PhysRevLett.31.1265}, in the rest of the paper we will focus only on the scalar case. The reason is twofold: from a practical point of view, the Klein-Gordon equation for a massive scalar field in a Kerr background is separable in Boyer-Lindquist 
coordinates~\cite{Carter:1968, PhysRevD.5.1913}, which makes the problem analytically tractable (see Appendix~\ref{sec: BH-Cloud System}). This is not the case for massive vector\footnote{See \cite{Frolov:2018ezx} for recent progress in the separability of the Proca field in Kerr backgrounds.}
 and tensor fields. From a theoretical point of view, ultralight scalar fields arise naturally as particle candidates in various scenarios for physics beyond the Standard Model. The Kerr-scalar system is therefore a particularly well-motivated system to be studied.

\subsection{Gravitational Atom} \label{sec: The Gravitational Atom}

The equation of motion of a massive scalar field $\Psi$ around a rotating BH is
\beq
\left(g^{ab} \nabla_a \nabla_b - \mu^2\right) \Psi (t, \textbf{r})= 0\, ,  \label{eqn:KG}
\eeq
where $g_{ab}$ is the Kerr metric and $\nabla_a$ is the associated covariant derivative. For (real) solutions of \eqref{eqn:KG} in the non-relativistic limit, it is convenient to consider the ansatz\footnote{Complex solutions may be described in terms of two real fields without loss of generality.}
\beq
\Psi (t, \textbf{r})= 
\frac{1}{\sqrt{2\mu}} \left[ \psi (t, \textbf{r})\, e^{-i \mu t} + \psi^*(t, \textbf{r}) \, e^{+ i \mu t}  \right] ,
\label{eqn:ansatz}
\eeq 
where $\psi(t, \textbf{r})$ is a complex scalar field which varies on a timescale that is much longer than $\mu^{-1}$. Substituting (\ref{eqn:ansatz}) into (\ref{eqn:KG}), and keeping only the leading contributions in $r^{-1}$, the field $\psi(t,\textbf{r})$ satisfies the Schr\"odinger equation with a Coulomb-like central potential\footnote{Although the Schr\"odinger equation is first order in time derivatives, the Klein-Gordon equation is second order. This means that generic solutions of \eqref{eqn:KG} will still depend on the value of the field and its first time derivative as initial conditions.}

\beq
i \frac{\partial }{\partial t} \psi(t, \textbf{r}) = \left[ - \frac{1}{2\mu} \nabla^2 - \frac{\alpha}{r} \right] \psi(t, \textbf{r}) \, , \label{eqn:Schrodinger}
\eeq
where we have defined the `fine-structure constant' 
\beq
\alpha \equiv \frac{G M \mu}{\hbar c} \simeq 0.02\left(\frac{M}{3 M_\odot}\right) \left(\frac{\mu}{10^{-12}\, \rm{eV}}\right)  . \label{eqn: fine-structure constant}
\eeq
Notice that $\alpha$ is the ratio of the gravitational radius of the BH, $r_g \equiv GM/c^2$, and the (reduced) Compton wavelength of the scalar field, $\lambda_c \equiv \hbar/(\mu c)$.

\subsubsection*{Energy spectrum}

  Remarkably,  the Schr\"odinger equation (\ref{eqn:Schrodinger}) takes the same form as for the hydrogen atom, such that the stationary eigenstates $\psi_{n \ell m}$ around the BH are given by the hydrogenic eigenfunctions (see Appendix~\ref{sec: BH-Cloud System} for details)
  \begin{align}
\psi_{n \ell m}(t, \textbf{r}) \simeq e^{- i (\omega - \mu) t} \bar{R}_{n \ell}(r) Y_{\ell m}(\theta, \phi) \, , \label{eqn: Hydrogen eigenstate}
\end{align}
where  $\{n,\ell,m\}$ are the principal, orbital, and magnetic `quantum numbers' respectively, which satisfy $\ell \le n-1$ and $|m| \le \ell$. For future convenience, we denote these eigenstates by $\psi_{n \ell m} \equiv \ket{n \ell m}$, and refer the occupied states as `clouds'. The eigenfrequencies are given by
\beq
\omega_{n \ell m} \simeq \mu \left(1- \frac{\alpha^2}{2n^2}\right) . \label{eqn: non-relativistic energy spectrum}
\eeq
The radial profile of the scalar cloud peaks at 
\beq r_{c, n} \simeq \left( \frac{n^2}{\alpha^2} \right) \,r_g \,,\eeq 
which can be far away from the central BH for $\alpha \ll 1$. 

\subsubsection*{Fine structure and beyond}	
\label{sec: fine structure and beyond}

\begin{figure}[t!]
\centering
\includegraphics[scale=1, trim = 0 0 0 0]{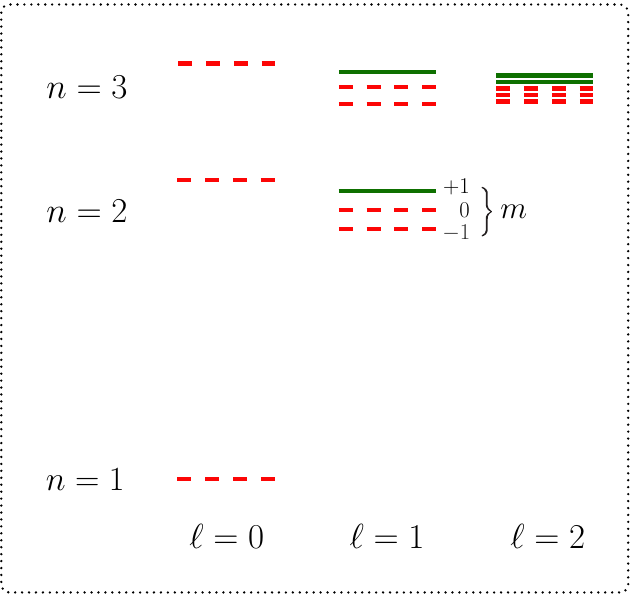}
\caption{Illustration of the eigenfrequency spectrum of the cloud, up to $n=3$ and $\l=2$. Green (solid) lines represent growing modes, while the red (dashed) lines are decaying modes. For a given value of $n$ and $\l$, the lowest frequency level is the $m=-\l$ mode, and subsequent higher frequencies are levels with larger values of $m$.} 
\label{fig:EnergyLevel2}
\end{figure}

Keeping higher powers of $r^{-1}$ in the large distance expansion, one obtains higher-order corrections to the eigenfrequencies (\ref{eqn: non-relativistic energy spectrum}) (see Appendix~\ref{sec: BH-Cloud System} for details): 
\beq
\Delta \omega_{n \ell m} = \mu \left(  - \frac{\alpha^4}{8 n^4} + \frac{ \left( 2\ell - 3 n + 1 \right) \alpha^4 }{n^4  (\ell + 1/2) }   + \frac{2 \tilde{a} m  \alpha^5}{n^3 \l (\l + 1/2) (\l + 1)}\right) .\label{eqn: change in frequency}
\eeq
The individual terms in (\ref{eqn: change in frequency}) are analogous to the relativistic correction to the kinetic energy, the fine splitting ($\Delta \l \neq 0$) and the hyperfine splitting ($\Delta m \neq 0$) in the hydrogen atom.
These corrections break the degeneracies between modes of the same $n$ but different $\l$ and $m$ (see~Fig.~\ref{fig:EnergyLevel2}).\footnote{Similar computations for the energy spectrum of bounded Dirac fields in Kerr spacetime can be found in \cite{Dolan:2015eua}.} As we shall see in Section~\ref{sec:perturbation}, the presence of an external perturbation can induce transitions between these energy levels. 

\subsubsection*{Decay width}

The analogy between the gravitational atom and the hydrogen atom is of course not exact. An obvious distinction is that electrons in the hydrogen atom are fermions, while the cloud is bosonic.  This means that  each level of the gravitational atom can be occupied by an exponentially large number of scalar quanta \cite{Arvanitaki:2010sy}, forming a boson {\it condensate}. 
Moreover, the states are not perfectly stationary, due to the boundary conditions at the horizon. This is encapsulated in the imaginary part of the eigenfrequency, \beq \omega_{n \ell m} \to \omega_{n \ell m} + i \Gamma_{n \ell m}\,,\eeq which was omitted in (\ref{eqn: non-relativistic energy spectrum}). In general, the instability rate has to be determined numerically~\cite{PhysRevD.76.084001}. However, in the limit $\alpha \ll 1$, the decay width can be computed analytically. The result, known as `Detweiler's approximation', is given by \cite{PhysRevD.22.2323} 
\beq
 \Gamma_{n \ell m} = \frac{2 r_+}{M} C_{ n \ell m }(\alpha)  \left( m \Omega_{\rm H} - \omega \right) \alpha^{4\ell + 5}\, ,  \label{eqn: Detweiler approximation}
\eeq
where we defined\hskip 1pt\footnote{See \cite{PhysRevD.86.104017, Rosa:2012uz} which point out a missing factor of $1/2$ in Detweiler's original computation. } 
\beq
C_{n \ell m}\left( \alpha \right) \equiv \frac{2^{4\ell +1} (n + \ell)!}{n^{2\ell + 4} (n - \ell- 1)!} \left[ \frac{\ell !}{(2\ell)! (2\ell + 1)!}\right]^2 \prod_{j=1}^{\ell} \left[ j^2 \left( 1 - \tilde{a}^2 \right)  + \left(\tilde{a} m - 2 \tilde{r}_+ \alpha  \right)^2 \right] , \label{eqn: Detweiler coeff}
\eeq
with $\tilde{a} \equiv a / M$ and $\tilde{r}_+ \equiv r_+ / M$.  Although the analytic result (\ref{eqn: Detweiler approximation}) was obtained for $\alpha \ll 1$, it was found  to give reasonable agreements with numerical results for $\alpha < 0.5$ (see Fig.~28 in~\cite{Brito:2015oca}).

\vskip 4pt
Due to the strong dependence of $\Gamma$ on $\ell$, the dominant growing modes have $\ell_g = m_g = 1$. In principle, this includes an infinite tower of overtones $\ket{ n_g 11}$, with $n_g \geq 3$, but, in practice, the fastest growing mode is $\ket{211}$. 
Moreover, at the end of the superradiant growth, we have $\omega_{n_g 11} > \omega_{211} = \Omega_{\rm H}$, so that the higher overtones decay back into the black hole. 
Hence, in the rest of the paper, we will consider $\ket{211}$ as the initial condition.  
The observationally relevant range of $\alpha$ is then given by
\beq
0.005 \left( \frac{M }{ 3 M_{\odot}} \right)^{1/9} \lesssim \ \alpha \ < 0.5 \, , \label{eqn: alpha constraint 1}
\eeq
where the lower bound is obtained by demanding the cloud to grow significantly within the age of the universe, while the upper bound is the maximum value of $\alpha$ for superradiant growth of the $\ket{211}$ mode to occur, assuming an initial BH spin of $\tilde{a} = 0.99$. Smaller values of $\tilde{a}$ result in stronger upper bounds on $\alpha$. For instance, $\tilde{a}=0.8$ implies $\alpha<0.25$~\cite{PhysRevD.76.084001}.

\subsection{Continuous Emission}	
\label{sec: Continuous Emission}

While the energy spectra are identical for both real or complex scalar fields, their stability properties may be different. The energy-momentum tensor of a real scalar field is always time dependent and non-axisymmetric. As a consequence, clouds made out of real fields are a continuous source of GWs~\cite{Arvanitaki:2010sy}. For complex fields, on the other hand, certain configurations may suppress GW emission. This occurs when the real and imaginary components of the field are in the same eigenstate, but their relative phase is such that they produce time-independent and axisymmetric configurations. Whether this is realized in a specific case depends on the initial conditions of the superradiant growth of the cloud. In any case, since the timescale for this GW emission is much longer than the timescale of the superradiance instability, the former doesn't inhibit the formation of the cloud~\cite{Yoshino:2014}. However, depending on the values of $M$ and $\alpha$, the clouds may still deplete on cosmological/astrophysical timescales. 

\vskip 4pt
The rate of GW emission for a real scalar cloud is given by\hskip 1pt\footnote{\label{footnote: continous decay}Comparison with numerical results~\cite{Yoshino:2014, Brito:2017zvb} suggests that (\ref{eqn: Schwarzschild strain 2}) is a good approximation for $\alpha \lesssim 0.1$. For larger values of $\alpha$, nonlinear effects reduce the emission power, and (\ref{eqn: Schwarzschild strain 2}) is only an upper bound.} 
\beq
\begin{aligned}
\frac{d E_{\rm gw}}{dt} \simeq 0.01 \left(\frac{M_c(\alpha)}{M} \right)^2 \alpha^{14} \, , \label{eqn: Schwarzschild strain 2}
\end{aligned}
\eeq
where the numerical coefficient was obtained from a fit to the numerical results of~\cite{Yoshino:2014, Brito:2017zvb} and $M_c(\alpha)$ is the mass of the cloud as a function of $\alpha$. For $\alpha \ll 1$, and assuming the initial BH spin to be close to maximal, $\tilde{a} \to 1$, we can approximate $M_c(\alpha) \simeq M \alpha$~\cite{Brito:2017zvb}.\footnote{For larger values of $\alpha$, determining $M_c(\alpha)$ requires numerical simulations; see~\cite{East:2017ovw} for the extraction efficiencies for axisymmetric complex Proca fields in near extremal BH backgrounds. The theoretical upper limit of superradiance extraction is given by $M_c/M < 0.29$ \cite{Christodoulou:1970}. } 
In the absence of additional effects such as accretion disks, conservation of energy implies $\dot{M}_c = - \dot{E}_{\rm gw}$, and the cloud evolves as
\beq
M_c(t) = \frac{M_{c, 0}}{1 + t/\tau_{c}} \,,  \label{eqn: cloud evolution}
\eeq
where $M_{c, 0}$ is the initial mass of the cloud and $\tau_c$ the lifetime of the cloud, which is given by
\beq
\begin{aligned}
\label{equ:tauc}
\tau_c &\,\simeq\,  10^7 \text{ years} \left( \frac{M}{3M_\odot} \right) \left( \frac{0.07}{\alpha}\right)^{15}\,,\\
&\,\simeq\,  
10^9 \text{ years} \left( \frac{M}{10^5 M_\odot} \right) \left( \frac{0.1}{\alpha}\right)^{15}   \, .
\end{aligned}
\eeq
These estimates are valid for $\alpha \lesssim 0.1$, while for larger values of $\alpha$, (\ref{equ:tauc}) can underestimate $\tau_c$ up to two orders of magnitude (see footnote \ref{footnote: continous decay}).

\vskip 4pt 
We see that the cloud's lifetime is extremely sensitive to the value of $\alpha$. For stellar-mass BHs, with masses in the range $[3,100]\,M_\odot$, we can ignore the GW emission from the cloud for typical binary dynamical timescales of order 10\,Myr to 1\,Gyr (e.g.~\cite{TheLIGOScientific:2016htt} and references therein), provided that $\alpha \lesssim 0.07$.   
On the other hand, for supermassive BHs with $M \gtrsim 10^5 M_\odot$, the cloud survives for more than 1\,Gyr if $\alpha \lesssim 0.1$. If the BHs continuously accrete matter, long-lived clouds may be allowed with larger values of $\alpha$~\cite{Brito:2014wla}. 
The associated ranges of $\alpha$ will become relevant when we discuss the evolution of the BH-cloud in a binary system, and in particular its GW signatures. 
While GW emission from the cloud may be ignored for evolution timescales that are smaller than~$\tau_c$, other depletion channels may become active in the presence of a companion. We will discuss this next.

\section{Clouds in Binary Systems}
\label{sec:perturbation}

So far, we have studied the BH-cloud in isolation, where it can be described by quasi-stationary states.  As we shall see, these configurations are altered when the BH carrying the cloud is part of a binary system (with a companion which may or may not carry a cloud itself). We will find that the gravitational perturbations on the BH-cloud yield interesting new phenomena. In particular, we will discover the presence of resonant orbital frequencies for the binary, which lead to new instabilities for the BH-cloud.  
We will discuss the observational consequences of these resonances in Section~\ref{sec:pheno}. 

\subsection{Gravitational Perturbations} \label{sec: gravitational perturbation}

The presence of a companion, of mass $M_*$, introduces a perturbation to the dynamics of the cloud. We will concentrate on the gravitational potential in the free-falling frame of the cloud, but also briefly discuss the possibility of mass/energy transfer between the cloud and the companion. As long as the binary separation, $R_*$, is larger than the size of the cloud, $R_* > r_c$,  the gravitational influence of the companion can be treated in a multipole expansion. Furthermore, we will assume that the correction to the Kerr metric is a small perturbation. This requires the length scale associated to the gravitational (curvature) perturbation $\sqrt{{R_*^3/M_*}}$ to be larger than $r_c$, or equivalently $R_* > 2^{-2/3} q^{1/3} \alpha^{2/3} r_c$, where $q \equiv M_* / M$. 
For $\alpha < 1$, the latter condition is somewhat weaker than the requirement of negligible mass transfer (see below), which reads $R_* > 2q^{1/3} r_c$. Taking this into account, the regime of validity of our perturbative treatment is bounded by $R_* > R_{\rm pt} \equiv \text{max} \{ r_c, \, 2 q^{1/3} r_c\}$.

\subsubsection*{Free-falling clouds}

The perturbed metrics for BHs in external fields have been studied extensively in the literature, see e.g.~\cite{Landry:2015zfa} and references therein.  The companion object induces a time-dependent perturbation to the metric, such that $g_{\mu\nu} = g^{(0)}_{\mu\nu} + \delta g_{\mu\nu}$, where $g^{(0)}_{\mu\nu}$ is the unperturbed Kerr spacetime. In general, the metric perturbation $\delta g_{\mu\nu}$ consists of two separate components: {\it i}\hskip 1pt) a direct contribution from the gravitational potential due to $M_*$, and {\it ii}\hskip 1pt)~the response of the BH-cloud to tidal deformations. We will work in the Newtonian limit, which dominates the gravitational perturbation to the cloud due to the companion, and ignore the subleading tidal contributions, as well as the effect of the BH spin on the metric. (We will discuss tidal and spin-induced effects in \S\ref{sec:binary}, when we study the GW signal from the binary system.)

\begin{figure}[t]
\centering
\includegraphics[scale=0.2, trim = 0 0 0 0]{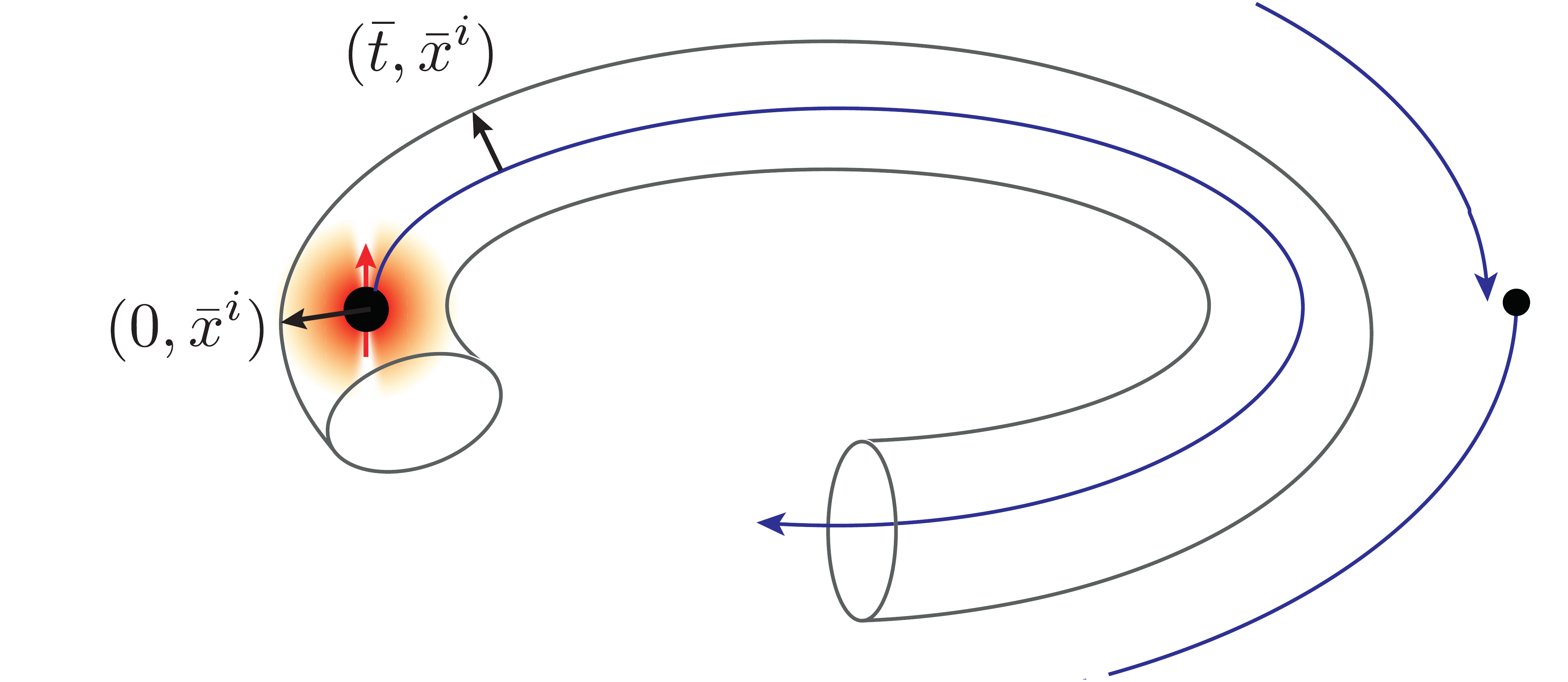}
\caption{Fermi coordinates $(\bar{t}, \bar{x}^i)$ centered on the geodesic of the BH-cloud system.} 
\label{fig: Fermi}
\end{figure}

\vskip 4pt
 In comoving Fermi coordinates $(\bar t, \bar x^i)$, whose origin is located at the centre-of-mass\hskip 1pt\footnote{In our case, the center-of-mass of the BH-cloud coincides with that of the isolated BH, such that the $(\bar t, \bar x^i)$ coordinates may be directly related to the Boyer-Lindquist coordinates of the Kerr metric.} of the unperturbed BH-cloud (see Fig.~\ref{fig: Fermi}), the $\delta g_{00}$ component reads~\cite{PhysRevD.34.991,Landry:2015zfa}  
\beq
 \begin{aligned}
\delta g_{00}(\bar t, \bar r) &\,=\,  \sum_{\ell_* =2} \frac{2}{\l_*(\l_* -1)}\, {\cal E}_L(\bar t, 0)\, \bar x^L  
\\
&\,=\, 
 \sum_{\ell_* =2} \sum_{|m_*| \leq \l_*} \frac{2}{\l_*(\l_* -1)} \,\bar r^{\l_*}\,  {\cal E}_{\l_*}^{(m_*)} (\bar t,0)\, Y_{\l_* m_*}(\bar \theta,\bar \phi) \, , \label{equ:harm}
 \end{aligned}
 \eeq
where we have used the notation $L \equiv (i_1\cdots i_{\l_*})$. The tensor ${\cal E}_L$ is the symmetric trace-free representation of the tidal tensor, defined as ${\cal E}_L (\bar{t}, 0) \equiv  -\partial_L U_*(\bar t, 0)/{(\l_*-2)!}$, with $U_*(\bar t)= -M_*/R_*(\bar t)$ the gravitational potential generated by the companion, evaluated along the geodesic of the BH-cloud.  
In the second line of \eqref{equ:harm} we have decomposed the gravitational field into spherical harmonics in polar coordinates $ \bar \r \equiv \{ \bar r,  \bar \theta,  \bar \phi\}$, with coefficients ${\cal E}_{\l_*}^{(m_*)} (\bar t,0)$.  
The fact that the expansion (\ref{equ:harm}) in the {\it free-falling} frame begins with two derivatives of the external potential is a well-known consequence of the equivalence principle. 

 \vskip 4pt 
The tidal tensors can easily be computed. For example, at leading Newtonian order, the $\ell_* =2$ mode is~\cite{Taylor:2008xy}
\beq
\begin{aligned}
\mathcal{E}_{ij}(t,0) \,\bar x^i \bar x^j & = - \frac{3 M_*}{R_*^3} \left( \hat n_i  \hat n_j - \frac{1}{3} \delta_{ij} \right) \bar x^i \bar x^j   \\
& =  -  \frac{8\pi}{5} \frac{M_*}{R_*} \left( \frac{\bar r}{R_*} \right)^2 \sum_{m_*=-2}^2 \  Y^*_{2 m_*} (\Theta_*, \Phi_*) Y_{2 m_*} (\bar \theta, \bar \phi) \,,
\label{eqn: Poisson tidal}
\end{aligned}
\eeq
where $\R_*(t)\equiv \{R_*(t),\Theta_*(t),\Phi_*(t)\}$ describes the position of the companion relative to the BH-cloud frame, $\hat {\bf n}\equiv \R_*/|\R_*|$, and $t$ now corresponds to the time measured by asymptotic observers, as opposed to observers in the free-falling frame.\footnote{The distinction between time coordinates only plays a role at higher post-Newtonian orders, e.g.~\cite{Landry:2015zfa}, when the boost factors begin to contribute.} At this order, we thus have
\beq
{\cal E}_{\ell_*=2}^{(m_*)}(t,0) = -  \frac{8\pi}{5} \frac{M_*}{R_*^3} Y^*_{2 m_*} (\Theta_*, \Phi_*)\,. 
\eeq
Using these results, and considering higher harmonics, it is straightforward to find the corrected gravitational dynamics in the non-relativistic limit. Substituting the perturbed metric into the Klein-Gordon equation, we find that, at lowest order, the gravitational potential in the Schr\"odinger-like equation \eqref{eqn:Schrodinger} is simply shifted by 
\begin{align}
V_*(t, \bar r) &\,=\, \frac{1}{2} \mu\hskip 1pt \delta g_{00} \,=\,  - \frac{M_* \mu}{R_*}\,  \sum_{\l_* \ge 2} \sum_{|m_*| \le \l_*} \frac{4\pi}{2\l_* + 1} \left(\frac{\bar r}{R_*}\right)^{\l_*}\,  Y^*_{\l_* m_*}(\Theta_*,\Phi_*) \,Y_{\l_* m_*}(\bar \theta,\bar \phi)\, \label{equ:dV} \, .
\end{align} 
Crucially, the first contribution starts with the {\it quadrupole} $\ell_*=2$. In principle, a different choice of observers/coordinates --- not free-falling with the BH-cloud  --- could lead to the appearance of extra terms in the potential, for instance a {\it dipole}. However, as we show in Appendix \ref{sec: Gravitational Perturbation Appendix},  this fictitious dipole eventually cancels.   
Finally, the above reasoning can be generalized to incorporate all types of relativistic corrections.
\vskip 4pt
In summary, the Klein-Gordon equation for the BH-cloud system receives corrections due to the presence of a companion which, in the non-relativistic limit, lead to a perturbed gravitational potential \eqref{equ:dV}. The salient feature is the fact that the multipolar interaction starts at the quadrupole level, as determined by the equivalence principle.

\subsubsection*{Mass transfer} \label{sec: Roche-Lobe}

\begin{figure}[t!]
\centering
\includegraphics[scale=0.15]{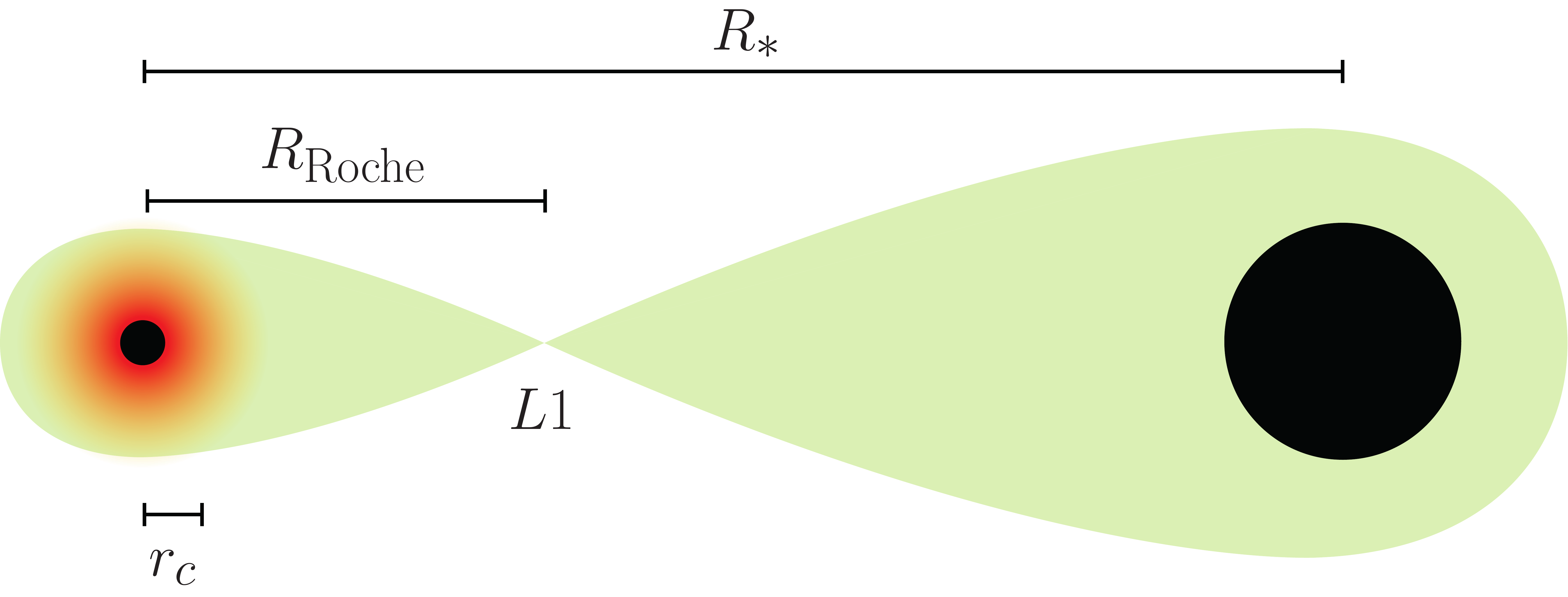}
\caption{Illustration of the Roche lobes for a binary with $q \gg 1$. As the separation decreases, the Roche lobe of the BH-cloud  begins to shrink. At the critical value $R_{*,{\rm cr}}$, given by \eqref{Rochemass}, the size of the cloud $r_c$ exceeds $R_{\rm Roche}$,
and mass transfer starts to occur.} 
\label{fig:Roche}
\end{figure}

The mutual gravitational attraction between the bodies can also induce transfer of mass/energy between the cloud and the companion. This happens when the characteristic Bohr radius~$r_c$ exceeds the Lagrange point, $L1$, located in between the two objects of the binary. The equipotential surface with the same gravitational potential as $L1$ is called the Roche lobe (see Fig.~\ref{fig:Roche}). 

\vskip 4pt 
Mass transfer from the cloud to the companion happens when $r_c \gtrsim R_{\rm Roche}$.
Using Eggleton's fitting formula~\cite{Eggleton:1983rx}, this can be converted into a critical orbital separation  
\beq
R_{*, \rm cr} \equiv   \left( \frac{0.49 \, q^{-2/3}}{0.6 \, q^{-2/3} + \ln \left( 1+q^{-1/3} \right)}  \right)^{-1} \, r_c \,.\label{Rochemass}
\eeq
This phenomenon becomes particularly important when $q \gg 1$, since the Roche lobe surrounding the BH carrying the cloud is then relatively small. In this limit, we find
\beq
R_{*,\rm cr} \simeq 2q^{1/3} r_c\,.\label{masst2}
\eeq 
In the rest of the paper, we will impose $R_{*} > R_{*,\rm cr}$, such that  mass transfer can be ignored.

\subsection{Level Mixing  }
\label{sec: level mixing}

The gravitational perturbation $V_*$ in \eqref{equ:dV} induces an overlap between the modes  $\ket{\psi_i} \equiv \ket{ n_i \l_i m_i}$ and $\ket{\psi_j} \equiv \ket{ n_j \l_j m_j}$, through $\bra{\psi_j} V_* \ket{\psi_i}$. Substituting (\ref{equ:dV}), we get 
\beq
\bra{\psi_j}  V_* \ket{\psi_i}=  - M_* \mu \sum_{\l_* \ge 2} \sum_{|m_*| \le \l_*} \frac{4\pi}{2\l_*+1} \frac{Y^*_{\l_*m_*}(\Theta_*,\Phi_*)}{R_*^{\l_*+1}} \times I_{\bar{r}} \times I_\Omega \, , \label{eqn: gravitational perturbation overlap}
\eeq
where
\begin{align}
I_{\bar{r}} &\equiv \int_0^\infty \d \bar{r} \, \bar{r}^{2+\l_*} \bar{R}_{n_j \l_j}(\bar{r}) \bar{R}_{n_i\l_i}(\bar{r}) \, , \\
I_{\bar{\Omega}} &\equiv \int \d \bar{\Omega} \ Y^*_{\l_j m_j}(\bar{\theta}, \bar{\phi})\, Y_{\l_i m_i}(\bar{\theta},\bar{\phi})\, Y_{\l_*m_*}(\bar{\theta},\bar{\phi})\, . 
\end{align}
The angular integral $I_{\bar{\Omega}}$ can be expressed in terms of the Wigner 3-j symbol,
\begin{align}
I_{\bar{\Omega}}
&= \sqrt{\frac{(2\l_j+1)(2\l_i+1)(2\l_*+1)}{4\pi}} \begin{pmatrix} \l_j & \l_i & \l_* \\ 0&0&0\end{pmatrix}  \begin{pmatrix} \l_j & \l_i & \l_* \\ -m_j&m_i& m_*\end{pmatrix} ,
\end{align}
which implies the following {\it selection rules}:
\begin{itemize}
\item[({\sf S1})]\ $-m_j+m_i+ m_*=0$\,,
\item[({\sf S2})]\ $|\l_j-\l_i| \le \l_* \le \l_i+\l_j$\,,
\item[({\sf S3})]\ $\l_i+\l_j+\l_* = 2p$, for $p\in \mathbb{Z}$\,.
\end{itemize}
Recall that the fastest growing mode $\ket{211}$ has $\l_g=m_g=1$, while the fastest decaying mode $\ket{100}$ has $\l_d=m_d=0$. In this case, the above selection rules would require $\l_*= \pm  m_*=1$, namely a {\it dipole} coupling, which is absent in \eqref{equ:dV}. For the quadrupole, the two fastest decaying modes that can couple to $\l_g=m_g=1$ are $\l_d =1, m_d=-1$ and $\l_d =1, m_d=0$. Since these rules are obtained purely from the angular dependence of the eigenfunctions, they apply equally to the fundamental mode ($ n=2$) and the overtones ($n \ge 3$). 

\subsection{Rabi Resonances} 
\label{sec: Time Evolution}

\begin{figure}[t!]
\centering
\includegraphics[scale=1., trim = 0 0 0 0]{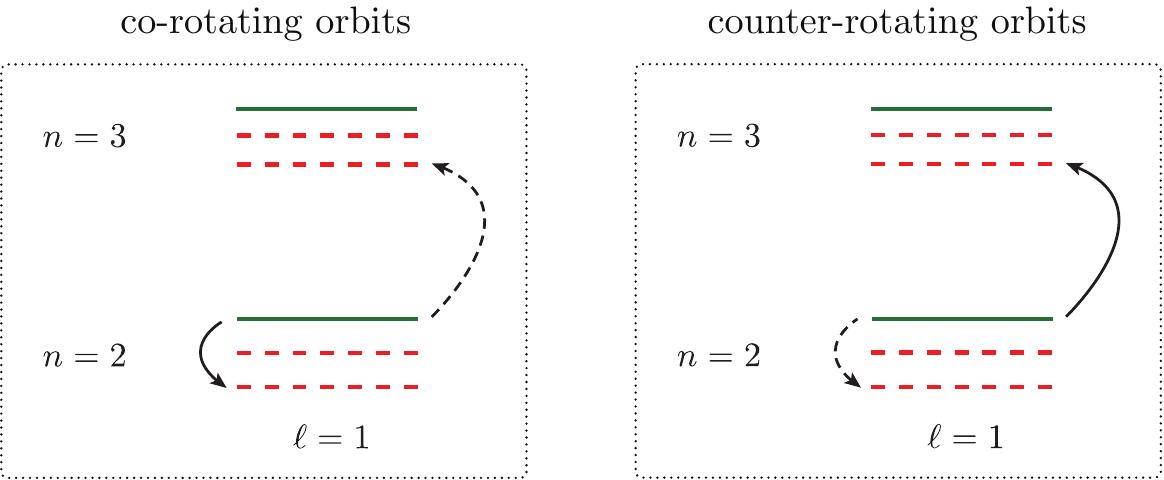}
\caption{Illustration of the frequency spectrum for $\ell=1$ up to $n = 3$ (there is in principle an infinite tower of overtones), for co-rotating and counter-rotating orbits. Solid arrows represent the allowed resonant transitions, while dashed arrows denote perturbative level mixings.}
\label{fig:EnergyLevelsOrbits}
\end{figure}

We now investigate how level mixings (see Fig.~\ref{fig:EnergyLevelsOrbits}), induced by the quadrupole $\l_*=2$, affect the dynamical evolution of the cloud. For simplicity, we will restrict ourselves to quasi-circular orbits with orbital frequency
\beq
\label{omega}
\Omega  \equiv \sqrt{\frac{M + M_*}{R_*^3}}\, ,
\eeq
where $M$ is the total mass of the BH-cloud, which  is equal to the initial BH mass {\it before} superradiance, if we neglect the small mass loss due to GW emission. Since $\Phi_*$ denotes the azimuthal angle of $M_*$ relative to the BH-cloud, 
 the orientation $\Phi_* = + \Omega t$, with $\Omega > 0$, corresponds to orbits that co-rotate with the cloud, called {\it co-rotating} orbits, while $\Phi_* = - \Omega t$ are {\it counter-rotating} orbits. This distinction will play a key role in the survival of the cloud throughout the inspiral stage. For notational convenience, we will write $\Phi_* = \pm \Omega t$, where the upper sign corresponds to co-rotating orbits and the lower sign denotes counter-rotating orbits. 

\subsubsection{Hyperfine Mixing} \label{sec: Hyperfine Mixing}

We first consider the mixing of the dominant growing mode $\ket{211}$ with the nearby decaying modes $\ket{210}$ and $\ket{21-1}$. The problem then involves the dynamics of a coupled three-state system. As we shall see, for orbital motions along the equatorial plane, the mode $\ket{210}$ decouples and the dynamics reduces to a two-state system.

\subsubsection*{Three-state system}  

For the three-state system, the expectation value of Hamiltonian for the field $\psi$ takes the form
\begin{align}
H = H_0  + H_1 + H_2\, ,
\end{align}
where $H_0$ represents the diagonal terms that contribute equally to all three modes, $H_1$ contains terms that  split the degeneracy, and $H_2$ are the off-diagonal terms induced by the quadrupole perturbation~$\ell_* = 2$. 
Using the basis $\ket{211} = (1, 0, 0)^{\text{T}}, \ket{210} = (0, 1, 0)^{\text{T}}$ and $\ket{21-1} = (0, 0,1)^{\text{T}}$,
the explicit matrix representations of the different contributions are
\beq
\begin{aligned}
H_0  & = \begin{pmatrix} E  & 0 & 0 & \\ 0 & E  & 0 \\ 0 & 0 & E \end{pmatrix} \, , \qquad 
H_1 = \begin{pmatrix} \epsilon_h + 3\eta f_0(\Theta_*)  & 0 & 0 & \\ 0 & - 6\eta f_0(\Theta_*) & 0 \\ 0 & 0 & - \epsilon_h + 3 \eta f_0(\Theta_*)  \end{pmatrix} \, , \\[4pt]
H_2 & =  \begin{pmatrix} 0  &  3 \eta f_1(\Theta_*) e^{\mp i \Omega t} & 3\eta  f_2(\Theta_*) e^{\mp 2 i  \Omega t}  & \\  3\eta f_1(\Theta_*) e^{\pm i \Omega t} & 0 & - 3\eta   f_1(\Theta_*) e^{\mp i \Omega t} \\  3\eta f_2(\Theta_*) e^{\pm 2 i  \Omega t}  & - 3 \eta  f_1(\Theta_*) e^{\pm i \Omega t} & 0 \end{pmatrix}  \, , \label{eqn: Hamiltonian matrices}
\end{aligned}
\eeq
where we have introduced the following quantities (see Appendix~\ref{sec: BH-Cloud System}) 
\begin{align}
E &\equiv \mu \left(- \frac{1}{8} \alpha^2 - \frac{17}{128} \alpha^4 \right)    ,  \label{eqn: diagonal E} \\
\epsilon_h & \equiv \frac{\mu}{12}\, \tilde{a}\, \alpha^5 \,   , \label{eqn: epsilon Hyperfine}  \\ 
\eta & \equiv  \, \alpha^{-3} \left( \frac{ q}{R_*} \right)   \left( \frac{ r_g }{R_*}\right)^2  .  
\end{align}
The parameter $\epsilon_h >0$ is the hyperfine splitting, and $\eta$ characterized the strength of the perturbation, proportional to the expectation value $\bra{\psi_j}  V_* \ket{\psi_i}$ given by (\ref{eqn: gravitational perturbation overlap}) (with $i, j = 1, 2, 3$). The oscillatory terms in $H_2$ are due to the phase evolution $\Phi_* = \pm \Omega t$. The functions $f_{|m_*|}(\Theta_*)$ arise from the angular dependence of the $\ell_* = 2, m_* = 0, \pm 1 $ and $\pm 2$ modes:
\beq
\begin{aligned}
f_0(\Theta_*) & \equiv 3 \cos^2 \Theta_* - 1 \,   ,  \\
f_1(\Theta_* ) & \equiv  3 \sqrt{2} \cos \Theta_* \sin \Theta_* \,  ,  \\
f_2(\Theta_*) & \equiv 3 \sin^2 \Theta_* \, .
\end{aligned}
\eeq
For the special case $\Theta_* = \pi/2$ (i.e.~equatorial motion), this reduces to $f_0 = -1, f_1 = 0, f_2 = 3$. 

\vskip 4pt
It is convenient to work in the interaction picture, where the evolution of the state $\ket{\psi_I}$ is given by \begin{align}
\ket{\psi_I(t)} = e^{ i H_I t}  \ket{\psi(t)}   ,
\end{align}
where $H_I \equiv H_0 + H_1$ is time-independent.\footnote{Strictly speaking, $H_I$ has a time dependence due to the decrease of the orbital radius $R_*(t)$ in the binary. However, since $\dot{R}_* / R_* \ll \Omega$ during the inspiral stage, this time dependence can be treated adiabatically, and neglected at leading order. } Since $H_I$ is diagonal, it does not couple modes of different angular momenta.  The eigenstates of $H_I$ are therefore equal to the corresponding eigenstates of the non-relativistic hydrogen atom, with the same $\ell$ and $m$, up to $\mathcal{O}(\alpha^2)$ corrections which will not be relevant in our case. 
\subsubsection*{Two-state system}

We now consider quasi-circular orbits along the equator, where $f_1=0$. This reduces the problem to a coupled two-state system, which can be solved {\it exactly}. Labelling the growing and decaying modes as $\ket{\psi_g} \equiv \ket{211}$ and $\ket{\psi^{(h)}_d} \equiv \ket{21-1}$, the state in the interaction picture 
is expressed as a linear combination 
\beq
\ket{\psi_I(t)} = c_g(t) \ket{\psi_g} + c^{(h)}_d(t) \ket{\psi^{(h)}_d}  , \label{eqn: linear combination}
\eeq
where the time-dependent coefficients satisfy
\beq
| c_g(t) |^2 + |c^{(h)}_d(t)|^2 = 1\, .
\eeq
The Schr\"odinger equation becomes 
\begin{align}
 i \frac{d}{dt}  \begin{pmatrix} c_g(t) \\[4pt] c^{(h)}_d(t)
\end{pmatrix}
=  \begin{pmatrix} 0  &  9 \eta \, e^{- 2 i  (\pm \Omega - \epsilon_h ) t }  &  \\  9 \eta \, e^{+ 2 i  ( \pm \Omega - \epsilon_h) t }  &  0 \end{pmatrix}  \begin{pmatrix} c_g(t) \\[4pt] c^{(h)}_d(t)
\end{pmatrix}  . \label{equ:Schrodinger}
\end{align}
Starting the evolution in the growing mode, $c_g(0) = 1, c^{(h)}_d(0)=0$, we find 
\begin{align}
c_g(t) & =  \frac{e^{ i (\epsilon_h \mp \Omega ) t}}{ 2 \Delta^{(h)}_{R}} \bigg[   \left(\Delta^{(h)}_{R} + \epsilon_h \mp \Omega \right) e^{- i \Delta^{(h)}_{R} t } + \left( \Delta^{(h)}_{R} - \epsilon_h \pm \Omega \right) e^{+ i \Delta^{(h)}_{R} t }  \bigg] \, , \\
c^{(h)}_d (t) & = \frac{9 \eta\, e^{ - i (\epsilon_h \mp \Omega ) t}}{2 \Delta^{(h)}_{R}}  \bigg[   e^{- i \Delta^{(h)}_{R} t } - e^{+ i \Delta^{(h)}_{R} t }   \bigg] \, ,  \label{eqn: cg and cd}
\end{align}
where we introduced 
\beq
 \Delta^{(h)}_{R} \equiv   \sqrt{ (9\eta)^2 + (\epsilon_h \mp \Omega)^2 } \, . \label{eqn: DeltaR }
\eeq 
Since $\psi$ is a complex scalar field with a global $U(1)$ symmetry, there is an associated conserved Noether current $J^\mu$ (see Appendix~\ref{sec: KG Norm} for details).  At leading order, $
J^0 \simeq \psi^* \psi$, which is interpreted as the occupation density of the system, and is analogous to the probability density in quantum mechanics. 
The occupation density of the decaying mode is proportional to
\beq
| c^{(h)}_d(t) |^2  = \left[ 1 -\left( \frac{\epsilon_h \mp  \Omega }{\Delta^{(h)}_{R} } \right)^2   \right] \sin^2 \left[ \int_{t_0}^t \d t^\prime \,\Delta^{(h)}_{R}(t^\prime)  \right]  . \label{eqn: modulus square of cg and cd 1} 
\eeq
Note that $\Delta^{(h)}_R$ is not constant during the inspiral, but increases as the orbit shrinks, so that the phase of the oscillations in (\ref{eqn: modulus square of cg and cd 1}) has been written as an integral over time. We see that $2\Delta^{(h)}_{R}$ controls the frequency of oscillations between the growing and decaying modes.  In analogy with the quantum mechanical problem, we call this the {\it Rabi frequency}. 

\subsubsection*{Hyperfine resonance}

When the orbital frequency $\Omega$ matches the hyperfine splitting $\epsilon_h$, the system experiences a resonance, and starts to oscillate between the growing and decaying modes. Since $\epsilon_h > 0$,  the resonance will only take place for co-rotating orbits. We will refer to this effect as the {\it hyperfine resonance}. This happens when the binary separation is
\beq
\begin{aligned}
R^{(h)}_{\text{res}} & = 144^{1/3} \, \alpha^{-4} (1+q)^{1/3} \hskip 1pt\tilde{a}^{-2/3} \, r_g \, , \\ 
& \simeq 9^{1/3} \, \alpha^{-14/3} \left( 1 + 4 \alpha^2 \right)^{2/3} (1+q )^{1/3} \left( \frac{a_s}{a}\right)^{2/3} \, r_g  \, , \label{eqn: Rres}
\end{aligned}
\eeq
where we have used (\ref{eqn: BH spin saturation}) for $a_s$, with $M_s \simeq M$ and $M_s \omega_s \simeq \alpha$. 

\subsubsection{Bohr Mixing}  

We now consider the possibility that the dominant growing mode $\ket{211}$ mixes also with the decaying modes $\ket{n 1 0}$ and $\ket{n 1 -1}, $ with $n \geq 3$. We will take the coupling to the $n=3$ modes to be a proxy for the mixing with this infinite tower of overtones. Similar conclusions hold for the higher-order overtones ($n \geq 4$). In particular, since $\omega_{n 1-1} > \omega_{31-1}$, the only difference is that these resonant mixings occur at orbital separations that are slightly shorter.  As before, we restrict ourselves to motion in the equatorial plane, $\Theta_* = \pi/2$, so that the $\ket{310}$ mode decouples. 
\subsubsection*{Three-level system}

\begin{figure}[t!]
\centering
\includegraphics[scale=1, trim = 0 0 0 0]{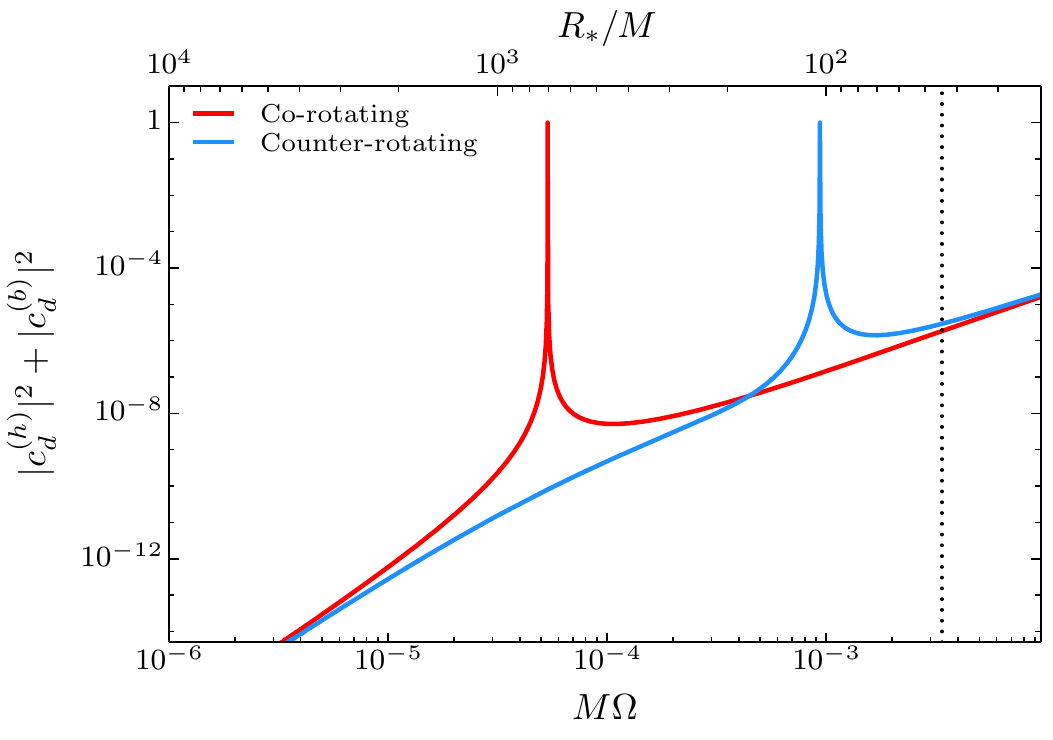}
\caption{ Evolution of the occupation density of the decaying modes, near the hyperfine and Bohr resonances, for co-rotating (red) and counter-rotating (blue) orbits, respectively. For illustration purposes, we use $\alpha=0.3$ and $q=10^{-3}$. The dotted vertical line denotes the characteristic Bohr radius of the $\ket{211}$ mode. For clarity, the oscillatory features of the solutions (\ref{eqn: modulus square of cg and cd 1}) and (\ref{eqn: modulus square of cg and cd 1 Bohr}) have been omitted. }\label{fig: Envelopes}
\end{figure}

 Labelling $\ket{\psi_g} \equiv \ket{211}$ and $\ket{\psi^{(h)}_d} \equiv \ket{21-1}$ as before, but adding the extra decaying mode $\ket{\psi^{(b)}_d} \equiv \ket{31-1}$, the state of the system in the interaction picture is now expressed as the following linear combination 
\beq
\ket{\psi_I(t)} = c_g(t) \ket{\psi_g} + c^{(h)}_d(t) \ket{\psi^{(h)}_d} +c^{(b)}_d(t) \ket{\psi^{(b)}_d} \, , \label{eqn: linear combination}
\eeq
with the normalization condition given by
\beq
|c_g|^2 + |c_d^{(h)}|^2 + |c_d^{(b)}|^2 = 1 \, . \label{eqn:normalization3}
\eeq
The Schr\"odinger equation implies 
\begin{align}
 i \frac{d}{dt}  \begin{pmatrix} c_g(t) \\[4pt] c_d^{(h)}(t) \\[4pt] c_d^{(b)}(t)
\end{pmatrix}
=  \begin{pmatrix} 0  & \ 9 \eta \, e^{- 2 i  (\pm \Omega - \epsilon_h ) t } \ & - 7.6\hskip 1pt \eta \, e^{- 2 i  (\pm \Omega - \epsilon_b ) t }  &  \\[4pt]  \ 9 \eta \, e^{+ 2 i  ( \pm \Omega - \epsilon_h) t }\   &  0 & 0 \\[4pt]
 \ - 7.6\hskip 1pt \eta \, e^{+ 2 i  ( \pm \Omega - \epsilon_b) t }  &  0 & 0
 \end{pmatrix}  \begin{pmatrix} c_g(t) \\[4pt] c_d^{(h)}(t) \\[4pt] c_d^{(b)}(t) 
\end{pmatrix}  , \label{equ:Schrodinger3}
\end{align}
with 
\beq
\epsilon_b  \equiv -  \frac{5 }{144}\,\mu\, \alpha^2 \,. 
\label{equ:EpsEta}
\eeq
\vskip 6pt
\noindent
In general, there is no analytic solution for the three-level system (\ref{equ:Schrodinger3}).  However, 
since we have $\epsilon_h/|\epsilon_b| \simeq \alpha^{3} \ll 1$, there is a clear hierarchy of scales in the evolution equation, 
and the system will probe different decaying modes at different times. The strength of the coupling --- away from resonances ---  is determined by the ratio of the size of the gravitational perturbation to the energy split. During the early stages of the inspiral, when $\Omega \lesssim \epsilon_h$, the mixing with the hyperfine state dominates. The solutions for $c_g(t)$ and $c_d^{(h)}(t)$ are thus identical to those given in~\eqref{eqn: modulus square of cg and cd 1}, with an error in the normalization condition of order $(\epsilon_h /\epsilon_b)^2$.  
As we have seen, for co-rotating orbits, the binary experiences a resonance when $\Omega = \epsilon_h$, while counter-rotating orbits continue smoothly through this region. When the orbital frequency approaches the scale of the Bohr splitting, $\Omega = |\epsilon_b|$, the overtone $\ket{31-1}$ gets excited, see Fig.~\ref{fig: Envelopes}.

\subsubsection*{Bohr resonance} 

Near $\Omega \simeq |\epsilon_b|$, the phase of the hyperfine mixing term in \eqref{equ:Schrodinger3} oscillates rapidly over a time of order $\eta^{-1}$. In this region, we have
 $|c_d^{(h)}|^2 \sim (\eta/\Omega)^2 \sim (\eta/\epsilon_b)^2 \ll 1$, so that we can ignore the small correction due to the $\ket{21-1}$  state. The dynamical evolution, once again, reduces to a two-level system, this time describing the mixing between the states $\ket{211}$ and $\ket{31-1}$. Repeating the analysis of \S\ref{sec: Hyperfine Mixing}, we find 
\beq
\begin{aligned}
 | c^{(b)}_d(t) |^2 & = \left[ 1 -\left( \frac{\epsilon_b \mp  \Omega }{\Delta^{(b)}_{R} } \right)^2   \right] \sin^2 \left[ \int_{t_0}^t \d t^\prime \,\Delta^{(b)}_{R}(t^\prime)  \right]  , \label{eqn: modulus square of cg and cd 1 Bohr} 
\end{aligned}
\eeq
where the modified Rabi frequency is
\beq
\Delta_R^{(b)} \equiv   \sqrt{ \left(7.6\hskip 1pt\eta\right)^2 + (\epsilon_b \mp \Omega)^2 } \, . \label{eqn: DeltaR Bohr }
\eeq 
Similar to the hyperfine case, the system undergoes a resonance when the orbital frequency matches with the energy split with the overtones. We refer to this effect as the {\it Bohr resonance}. A crucial difference to the case discussed in \S\ref{sec: Hyperfine Mixing} is that, while for the hyperfine resonance we have $\omega_{21-1} < \omega_{211}$,  the Bohr resonance is for an excited state $\omega_{31-1}> \omega_{211}$; see 
Fig.~\ref{fig:EnergyLevelsOrbits}. This implies that the Bohr resonance can only occur for counter-rotating orbits, at a  binary separation given~by 
\begin{align}
R^{(b)}_{\text{res}} = \left( \frac{144}{5} \right)^{2/3} \alpha^{-2} (1+q)^{1/3} \, r_g\, . \label{eqn: RresB}
\end{align}
Note the $\alpha^{-2}$ scaling, whereas the corresponding result for the hyperfine resonances scales as~$\alpha^{-4}$. For small~$\alpha$, the Bohr resonance therefore occurs at a much smaller separation than the hyperfine case (see Fig.~\ref{fig: Envelopes}). In terms of the Bohr radius of the cloud, we can write $R^{(b)}_{\rm res} =(18/5)^{2/3} (1+q)^{1/3} \,r_c$. For $q < 1$, the Bohr resonance thus occurs at the boundary of the regime of validity of the multipole expansion, $R_* > r_c$.  On the other hand, for $q \gg 1$, the multipole expansion remains accurate, while the stronger constraint typically comes from the requirement for the validity of the perturbative treatment of the gravitational interaction with negligible mass transfer, corresponding to
 $R_* > R_{\rm pt} =2 q^{1/3} r_c$.

\subsection{Cloud Depletion}
\label{sec: Cloud Depletion}

So far, we have taken the growing and decaying modes to be stationary, while in 
reality both 
have non-zero decay widths, see \eqref{eqn: Detweiler approximation}. Although $\Gamma_{211}$ vanishes when the superradiant growth stops, the rates $\Gamma_{ n \ell m} $, with $m\le 0$, remain finite throughout the evolution of the BH-cloud. Since the gravitational perturbation of the companion can induce transitions to the decaying modes, this opens up a new depletion channel for the cloud. In this section, we will explore this effect in more detail, both for co-rotating and counter-rotating orbits. 

\vskip 4pt
As we shall see, it is convenient to express the time evolution of the system in terms of the orbital frequency, $\Omega$, of the binary.  For quasi-circular orbits, the relation between $t$ and $\Omega$ due to the leading quadrupolar GW emission is~\cite{Peters:1964zz}
\beq
t(\Omega) = \tau_0 \left[ 1 -  \left( \frac{\Omega_0}{\Omega}\right)^{8/3} \right]   , \label{eqn: t-R LO quadrupole}
\eeq
 where $\tau_0$ is the time to merger for an initial orbital frequency $\Omega_0$:
\beq
\frac{\tau_0}{M_{\rm tot}} \equiv \frac{5}{256} \frac{1}{ \nu} \left( \frac{1}{ M_{\rm tot} \Omega_0 }  \right)^{8/3}  \, , \label{eqn: time to merger}
\eeq
with $M_{\rm tot} \equiv M + M_*$ and $\nu \equiv M M_* / M_{\rm tot}^2$.  For simplicity,  we will assume that we can neglect the GW emission from the cloud throughout the coalescence of the binary.  For the case of real scalar fields, this requires the time to merger,  
$\tau_0$, to be shorter than the lifetime of the cloud, $\tau_c$, introduced in \S\ref{sec: Continuous Emission}. 
This constraint also applies to complex fields, except for the special case where the resulting configuration is time-independent and axisymmetric. 
Imposing the condition $\tau_0< \tau_c$ leads to 
\beq
\frac{\tau_0}{\tau_c} \simeq \frac{10^{-21}}{(M\Omega_0)^{8/3}} \, \frac{\left(1+q \right)^{1/3}}{q} \left(\frac{\alpha}{0.07}\right)^{15} < 1\, , \label{eqn: ttm vs lifetime}
\eeq
which can be translated into a bound on the initial orbital frequency
\beq
\begin{aligned}
\Omega_0  \gtrsim 0.9 \text{ mHz} \left( \frac{3M_\odot}{M} \right) \, \frac{(1+q)^{1/8}}{q^{3/8}} \left(\frac{\alpha}{0.07}\right)^{45/8} \equiv  \Omega_{\rm cr}\, . \label{eqn: critical orbital frequency}
\end{aligned}
\eeq
In the following, we will assume $\Omega > \Omega_{\rm cr}$ during the entirety of the binary inspiral. In~principle, we can also study binaries which form at smaller initial orbital frequencies, so that GW emission depletes the cloud before merger. We will return to this point in \S\ref{sec: Phenomenology}.

\subsubsection*{Level mixings}

We take  $t_0$ to be the time at which the superradiance has saturated.
To estimate the amount of depletion between times $t_0$ and $t$, we introduce the following quantity
\beq
\mathcal{A} (t, t_0) \equiv  \sum_{{n},\ell} \sum_{m\leq 0} |\Gamma_{n \ell m} | \int_{t_0}^{t} \d t^\prime \, \big| c_{n \ell m} (t^\prime) \big|^2\, , \label{eqn: estimator definition 0}
\eeq
where $c_{n \ell m}(t)$ is the overlap of the state $\ket{\psi}$ with the decaying modes $\ket{n \ell m}$, with $m\leq 0$. The `depletion estimator' (\ref{eqn: estimator definition 0}) is thus the ratio of the integrated time spent by the system in the decaying modes to the decay timescale $|\Gamma_{n \ell m} |^{-1}$, weighted by the occupation density of each state.  Physical quantities of the cloud, such as its mass and angular momentum, would decay as $ e^{-2\mathcal{A}(t, t_0)}$, where the factor of 2 in the exponential arises because the energy-momentum tensor depends quadratically on the field. 
In the limit where mixing with the decaying states is negligible, $\mathcal{A} \to 0$ and the cloud remains stable. On the other hand, significant depletion occurs when ${\cal A}$ becomes of order one. 
\vskip 4pt
For $\alpha<1$, the decay rate \eqref{eqn: Detweiler approximation} is suppressed for higher $\ell$'s, so that 
the sum in (\ref{eqn: estimator definition 0}) will be dominated by the modes with $\ell=1$ and $m=-1$ (the $m=0$ mode decouples for orbits along the equatorial plane). Restricting further to the dominant decaying channels, with $n_h = 2$ and $n_b = 3$,\footnote{Ignoring the $n \geq 4$ overtones amounts to underestimating the total contribution from $n \ge 3$ by approximately a factor of $2$, mostly from 
the evolution of the cloud prior to the resonances. This arises from the suppression of $\Gamma_{ n \ell m}$ for increasing $n$, see (\ref{eqn: Detweiler coeff}).} and using (\ref{eqn: t-R LO quadrupole}) to convert the integral over time in (\ref{eqn: estimator definition 0}) into an integral over orbital frequency, we have\footnote{In the numerical integration of (\ref{eqn: estimator in R}), we ignore the oscillatory terms in (\ref{eqn: modulus square of cg and cd 1}) and (\ref{eqn: modulus square of cg and cd 1 Bohr}), which overestimates $\mathcal{A}$ by roughly a factor of $2$.}
\beq
\mathcal{A} (\Omega, \Omega_0) \simeq \sum_{i = \{h,b\}} |\Gamma^{(i)}_d| \left[ \frac{5}{96} \frac{1}{\nu} \frac{1}{M_{\rm tot}^{5/3}} \int^{\Omega}_{\Omega_0}  \d \Omega^\prime \,   \Omega^{\prime \, -11/3} \,\big| c^{(i)}_d( \Omega^\prime) \big|^2  \right] \label{eqn: estimator in R}.
\eeq
Using Detweiler's approximation~(\ref{eqn: Detweiler approximation}), the relevant decay rates  are
\beq
\begin{aligned}
 |\Gamma^{(i)}_{d} | & =  \frac{B_{(i)}}{24} \frac{ \alpha^{10}}{M}  \, \left( \frac{1-4\alpha^2}{1+4\alpha^2} \right)^{2}  \left( \frac{2 }{1+4\alpha^2  }  + \tilde{r}_+ \right)   , \label{eqn: Detweiler spin saturation}
\end{aligned}
\eeq
where  $B_{(h)} \equiv 1$ and $B_{(b)} \equiv 256/729$, and we used  $\tilde a=\tilde a_s$, at saturation.  In the limit of small $\alpha$, the corresponding decay time is 
\beq
\begin{aligned}
\tau_d^{(i)} &\simeq 1 \,B_{(i)}^{-1} \text{ years} \left( \frac{M}{3M_\odot} \right) \left(\frac{0.07}{\alpha}\right)^{10} \, , \\
&\simeq 0.9 \,B_{(i)}^{-1} \text{ years} \left( \frac{M}{10^5 M_\odot} \right) \left(\frac{0.2}{\alpha}\right)^{10} \, . \label{eqn: resonance decay timescale}
\end{aligned}
\eeq
This timescale is very sensitive to $\alpha$, but can be shorter than the duration of GW observations, so that the decay can, in principle, occur in observational bands. In the following, we will use the estimator (\ref{eqn: estimator in R}) to analyze the stability of the BH-cloud against level mixing, during the inspiral phase of the binary dynamics.

\begin{figure}[t]
\centering
\includegraphics[scale=0.7, trim = 0 0 0 0]{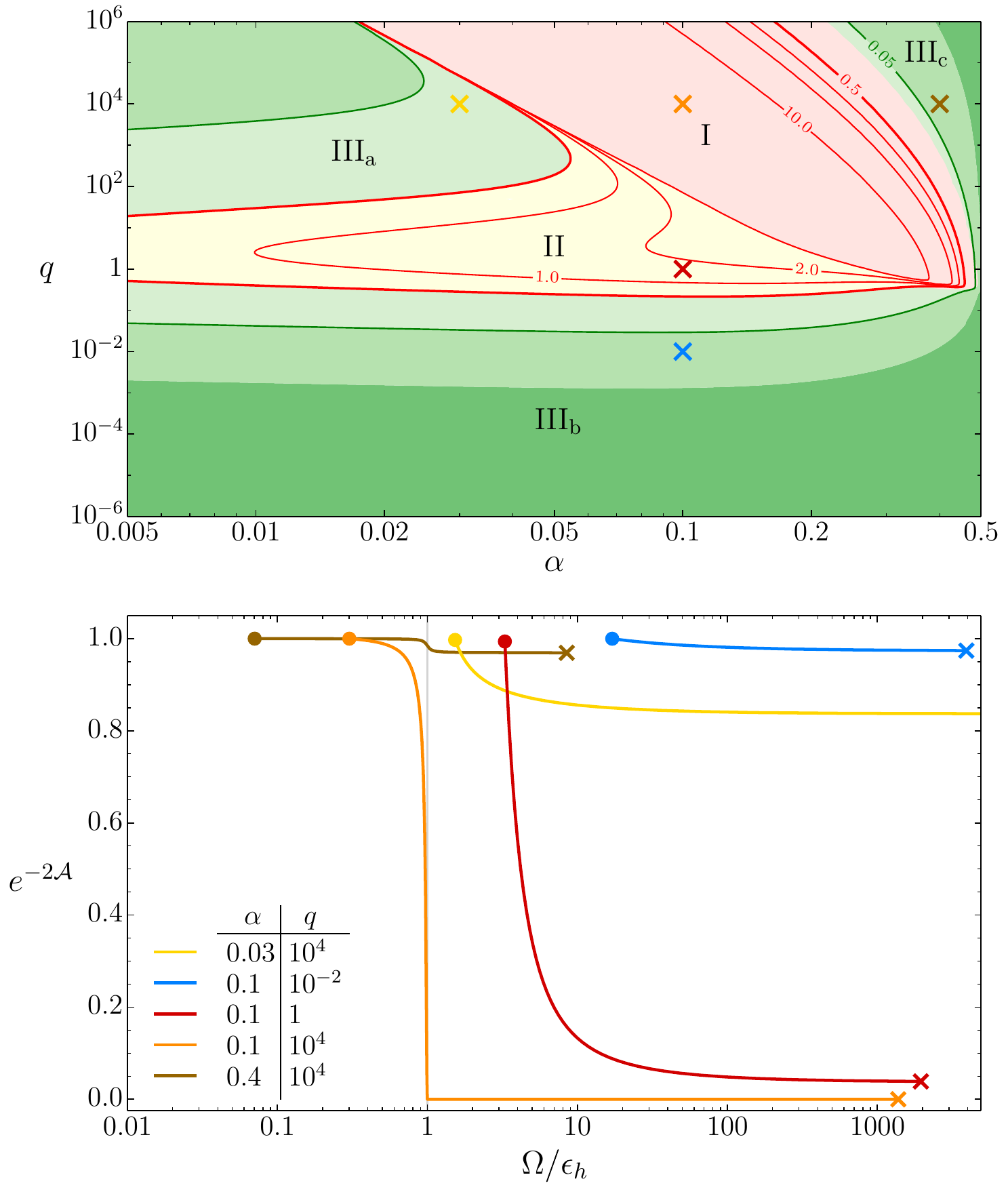}
\caption{Plot of the total depletion of the cloud for co-rotating orbits. Shown is ${\cal A}(\Omega_{\rm pt}, \Omega_{\rm cr})$, as a function of $q$ and $\alpha$ ({\it top}), and the evolution of $e^{-2{\cal A}}$ for specific choices of $q$ and $\alpha$ ({\it bottom}).  See the main text for further discussion.}\label{fig:co-rotating}
\end{figure}

\subsubsection*{Co-rotating orbits}
\label{sec: co-rotating orbits}

We first consider the evolution of the cloud for co-rotating orbits, which is dominated by the existence of the hyperfine resonance.  
For each combination of $\alpha$ and $q$, we chose the initial condition $\Omega_0 = \Omega_{\rm cr}$, see \eqref{eqn: critical orbital frequency}. 
Figure~\ref{fig:co-rotating} illustrates the total amount of depletion due to level mixing between $\Omega_0$ and the breakdown of perturbation theory at~$\Omega_{\rm pt}$. 
As shown in the top panel of Fig.~\ref{fig:co-rotating}, the parameter space divides into three distinct regions: 

\vskip 6pt
$\bullet$ Region~${\rm I}$ (red). \hskip 18pt The cloud depletes dramatically during the resonance.

$\bullet$ Region~{\rm II} (yellow). The cloud experiences a long period of perturbative depletion. 

$\bullet$ Region~{\rm III} (green). The cloud mostly survives the entire inspiral.

\vskip 6pt
\noindent
To understand the morphology of the parameter space, it is useful to compare the initial orbital frequency 
to that of the hyperfine resonance: 
\beq
\frac{\Omega_{0}}{\epsilon_h} \simeq 2.9 \left(\frac{0.1}{\alpha}\right)^{11/8} \frac{(1+q)^{1/8}}{q^{3/8}}\, .
\eeq
Only for $\Omega_{0} < \epsilon_h$ is the hyperfine resonance experienced during the part of the evolution shown in Fig.~\ref{fig:co-rotating}.
This occurs for $q \gtrsim 2\times10^{-4} \,\alpha^{-11/2}$, corresponding to regions ${\rm I}$ and ${\rm III}_c$ of the parameter space. In the other regions, the initial orbit has already passed the orbital frequency of the hyperfine resonance. 

\vskip 4pt
As seen in the example $\{\alpha=0.1,q=10^4\}$ (\text{\it orange curve}), for combinations of parameters in region~${\rm I}$, the cloud depletes dramatically during the hyperfine resonance. In region ${\rm III}_c$, on the other hand, the cloud survives the resonance, as  illustrated for the case $\{\alpha=0.4,q=10^4\}$ (\text{\it brown curve}). This is because, for large $q$ and $\alpha$, the binary only spends a short amount of time within the resonance epoch, relative to the decay time $\tau_d^{(h)}$.\footnote{For fixed mass of the BH-cloud, $M$, large values of $q$ increase the rate of GW emission and hence the shrinking of the orbit, while larger values of $\alpha$ push the resonance towards shorter distances. Notice that, while the GW emission is invariant under $q \to 1/q$, the timescale associated with the decaying mode, as well as the separation at resonance, is controlled by $M$, which breaks this symmetry.}

\vskip 4pt
The results in the other regions 
can be understood qualitatively as follows. The larger the value of $q$, for fixed $M$ and $\alpha$, the faster the binary transits through the inspiral stage, due to the efficiency of GW emission from the binary. At the same time, in the regime of perturbative level mixing, $|c_d|^2$ is proportional to $q$. Since $\mathcal{A}$ is roughly proportional to both the time spent in the inspiral and the strength of the coupling to the decaying modes, the two effects compete with each other. By continuity, $\mathcal{A}$ peaks at intermediate values of $q$ in region~$\rm{II}$.  As shown for the case $\{\alpha=0.1,q=1\}$ (\text{\it red curve}), the depletion can be significant, yet it extends over a larger time than in the case of the resonant decay. In regions~${\rm III}_a$ and~${\rm III}_b$, the rapid decay of the orbit due to GW emission ($q \gg 1$) and the suppression of the perturbative mixing $(q \ll 1)$ dominate, respectively. As a consequence, the cloud hardly depletes during the inspiral, as shown for $\{\alpha=0.03,q=10^4\}$ (\text{\it yellow curve}) and $\{\alpha=0.1,q=10^{-2}\}$ (\text{\it blue curve}). However, this does not necessarily imply that the cloud is stable beyond $\Omega > \Omega_{\rm pt}$, since our perturbative treatment of the problem starts breaking down.  Only a full numerical simulation could then inform us about the fate of the cloud towards the merger stage.

\begin{figure}[t]
\centering
\includegraphics[scale=0.7, trim = 0 0 0 0]{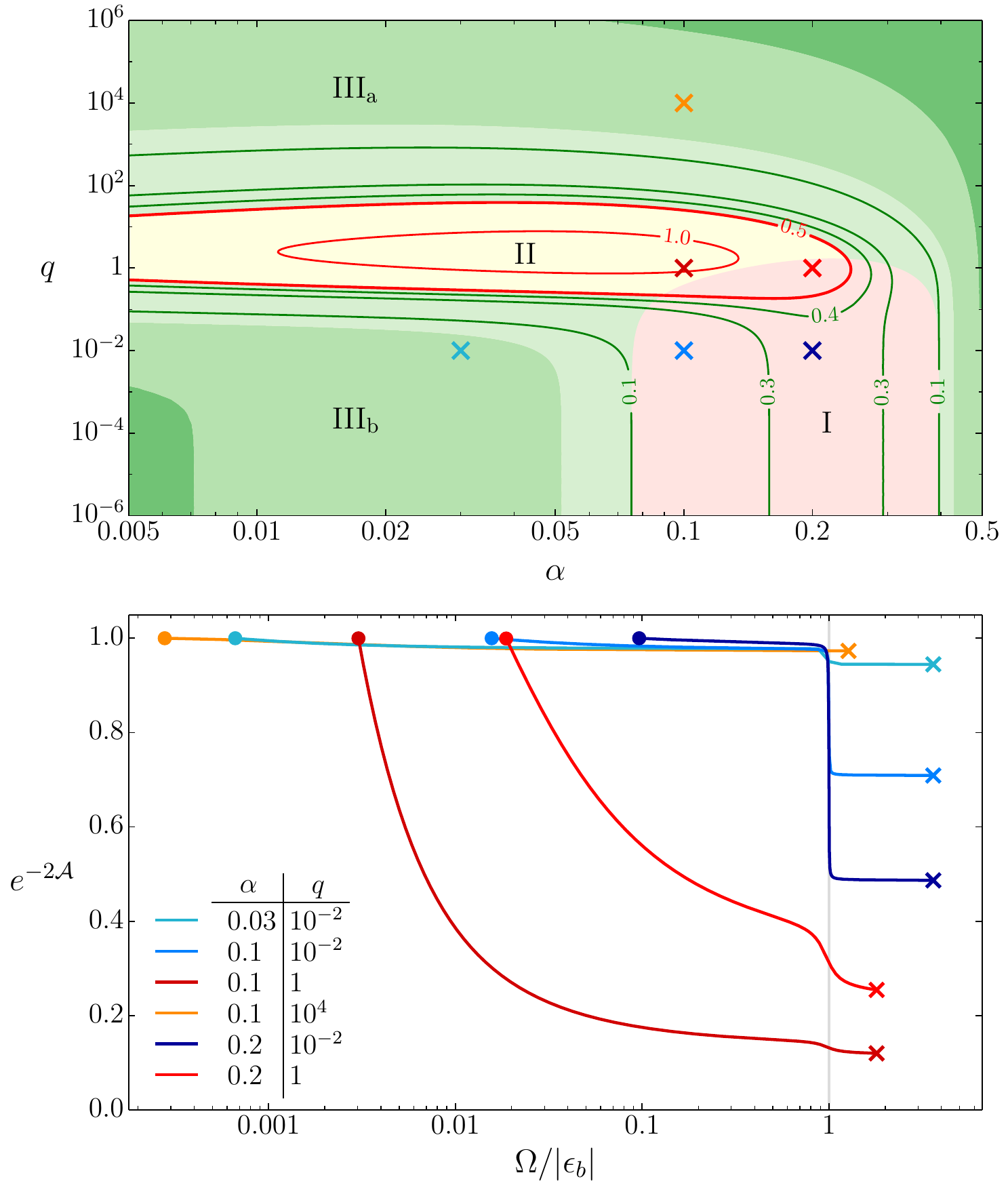}
\caption{Same as Fig.~\ref{fig:co-rotating}, but for counter-rotating orbits. 
The resonance is less pronounced than for co-rotating orbits since the system moves through it much faster.}
\label{fig:counter-rotating}
\end{figure}

\subsubsection*{Counter-rotating orbits}

Next, we consider the case of counter-rotating orbits, for which the Bohr resonance can be excited. 
As before, we chose the initial condition $\Omega_0 = \Omega_{\rm cr}$. Figure~\ref{fig:counter-rotating} illustrates the total amount of depletion between $\Omega_0$ and~$\Omega_{\rm pt}$. The labelling of the different regions is in accordance to Fig.~\ref{fig:co-rotating}. Once again, the basic features can be understood as follows.

\vskip 4pt
The ratio of the initial orbital frequency and the frequency of the Bohr resonance is given by:
\beq
\frac{\Omega_{0}}{\epsilon_b} \simeq  0.003 \, \left( \frac{\alpha}{0.1} \right)^{21/8}\frac{(1+q)^{1/8}}{q^{3/8}} \,.
\eeq
Except for very large values of $\alpha$ and very small $q$, this implies $\Omega_0 < \epsilon_b$, so that essentially all the orbits shown in Fig.~\ref{fig:counter-rotating} experience the Bohr resonance at some point during the inspiral. However, since the Bohr resonance occurs at much smaller orbital separations than the hyperfine resonance, the binary moves through it much faster. As a consequence, the amount of depletion 
is significantly less dramatic than for co-rotating orbits experiencing the hyperfine decay. Nevertheless, for $\alpha \gtrsim 0.07$ and $q \lesssim 1$ (corresponding to region I in the figure), the Bohr resonance still creates a sharp depletion of the cloud, yet without termination. This is illustrated by the {\it blue curves} in the bottom panel of Fig.~\ref{fig:counter-rotating}. 

\vskip 4pt
In region II, the cloud experiences a slow perturbative depletion---mostly controlled by mixing with the nearby hyperfine states---which can significantly reduce its energy density; see the {\it red curves} in Fig.~\ref{fig:counter-rotating}. By the time the cloud reaches the Bohr resonance, the depletion (in comparison) is only a small effect. In regions~${\rm III}_a$ and ${\rm III}_b$, the cloud survives with hardly any depletion, even after moving through the resonance. 

\newpage
\section{Gravitational Wave Signatures}
\label{sec:pheno}

In this section, we will briefly sketch some of the observational consequences of our findings. 
For real scalar fields, there are two sources of GWs: {\it i}\hskip 1pt) the continuous emission from the rotating cloud itself (\S\ref{sec:cloud}), and {\it ii}\hskip 1pt) the radiation produced by the binary (\S\ref{sec:binary}). 
Depending on the initial conditions, only the latter may be present for complex fields in axisymmetric configurations.
These signals can be affected by the dynamics of the boson cloud as the system evolves during the inspiral.  In \S\ref{sec: Phenomenology}, we will discuss a few paradigmatic examples of binary systems that can provide key information about the ultralight particles forming the cloud, and whose GW signatures are within reach of current and planned GW observatories.  Our treatment will be rather schematic and a more detailed investigation will be the subject of future work.

\subsection{Signal from the Cloud}
\label{sec:cloud}

As we have seen in \S\ref{sec: Continuous Emission}, a real scalar cloud emits continuous, monochromatic GWs~\cite{Arvanitaki:2010sy, Yoshino:2014}. The frequency of these GWs is determined by the mass of the scalar field,
\begin{align}
f_{c} \simeq \frac{\mu}{\pi} = 484 \, \text{Hz} \, \left(\frac{\mu}{10^{-12} \text{ eV}} \right) , \label{eqn: GW annihilation frequency} 
\end{align}
and their rms strain amplitude is~\cite{Brito:2014wla, Yoshino:2014} 
\begin{align}
h_c \simeq 2 \times 10^{-26} \left( \frac{M}{3 M_{\odot}} \right) \left( \frac{M_c(\alpha)/M}{0.1} \right) \left( \frac{\alpha}{0.07 }\right)^6  \left( \frac{10 \,  \text{kpc}}{d} \right), \label{eqn: GW annihilation strain 1}
\end{align}
where $d$ is the (Euclidean) distance of the source. 
Since the signal is emitted continuously, its detectability depends on the total observation time $T_{\rm obs}$. The signal-to-noise ratio (SNR) of the cloud signal is
\begin{align}
\text{SNR} = h_c \, \langle F^2_+ \rangle^{1/2}  \frac{T_{\rm obs}^{1/2}}{\sqrt{S_n(f_c)} } \, ,
\end{align}
where $S_n(f_c)$ is the (one-sided) noise spectral density at the frequency $f_c$, and $\langle F^2_+ \rangle = \langle F^2_\times \rangle$ is the angular average 
 of the square of the detector pattern functions $F_{+, \times}$ for each GW polarization.   
Using  
(\ref{eqn: GW annihilation strain 1}),  we get
\beq
\text{SNR} \simeq   13 \, \langle F^2_+ \rangle^{1/2} \left( \frac{T_{\rm obs}}{1 \text{ yr}} \right)^{1/2} \left( \frac{10^{-23} \text{ Hz}^{-\frac{1}{2}}}{\sqrt{S_n(f_c)}} \right) \left( \frac{M}{3 M_{\odot}} \right)  \left( \frac{M_c/M}{0.1} \right) \left(\frac{\alpha}{0.07} \right)^6 \left( \frac{10 \,  \text{kpc}}{d} \right) ,  \label{eqn: SNR  annihilation} 
\eeq
which is strongly dependent on $\alpha$. 
We see that BH-cloud systems with $\alpha < 0.07$ and $M < 10^2\, M_\odot$ may be detectable only within our own galaxy, while those with larger values of $\alpha$, and bigger masses, may be observed at extra-galactic distances. 
Further discussion and detailed forecasts can be found in~\cite{Arvanitaki:2014wva, Arvanitaki:2016qwi, Brito:2017zvb,Riles:2017evm}.

\subsubsection*{Resonant extinction/attenuation} 

When the BH-cloud is part of a binary system, the cloud may deplete over time, according to:\footnote{The ratio (\ref{eqn: cloud mass ratio}) need not be a smooth exponential decay, but 
may contain periodic features; cf.~\eqref{eqn: modulus square of cg and cd 1} and (\ref{eqn: modulus square of cg and cd 1 Bohr}). The depletion process switches on and off as the binary transits through the resonance. This effect manifests itself most prominently during the Bohr resonance.} 
\beq
 \frac{h_c(t)}{h_c(t_0)} = \frac{M_c(t)}{M_c(t_0)} \sim e^{- 2\mathcal{A}(t, t_0)} \, , \label{eqn: cloud mass ratio}
\eeq
where $\mathcal{A}$ is the depletion estimator introduced in (\ref{eqn: estimator definition 0}). 
This decay is most prominent for co-rotating orbits, since the hyperfine resonance occurs at large separations, so that the binary spends a significant amount of time near the resonance orbit. For counter-rotating orbits, the Bohr resonance attenuates the signal, but does not completely terminate it.
In either case, the extinction/attenuation of the GWs produced by the cloud turns into a unique feature of the binary system.  
In \S\ref{sec: Phenomenology}, we will discuss the phenomenological consequences of this effect in several case studies.

\subsubsection*{Doppler modulation}

In the presence of a companion, the GW frequency $f_c$ will also be modulated by the orbital motion.\footnote{This is the same effect that famously led to the discovery of the Hulse-Taylor binary pulsar, where the frequency of the emitted electromagnetic radiation was found to be modulated by the orbital motion~\cite{Hulse:1974eb}.} In particular, if the orbital plane of the binary is oriented such that the BH-cloud has a non-vanishing radial velocity along the line-of-sight, the orbital motion will induce a periodic Doppler shift of the GW signal,
\beq
\frac{\Delta f_c}{f_c} = \frac{v_{r}}{c} = \left( \frac{M_* }{M_{\rm tot}} \right) \frac{R_* \Omega }{c} \sin \iota \, ,
\eeq
where $v_{r}$ is the radial component of the velocity of the BH-cloud along the line-of-sight, and $\iota$ is its angle relative to the normal of the orbit. While detecting the periodic modulation of the frequency is experimentally (and computationally) challenging, this Doppler effect would be very convincing evidence that the BH-cloud is part of a binary system. At the same time, it will open a new avenue to infering parameters of a binary system at low orbital frequencies, by monitoring the signal from the cloud with continuous searches. Similar searches are being performed for neutron stars in binaries~\cite{Aasi:2014erp}.

\subsection{Signal from the Binary}
\label{sec:binary}

The presence of the cloud can also reveal itself in the GW signal of the binary, through subtle modifications in the waveforms due to the cloud's multipole moments and the tidal response to the companion. In the following, we will show how these finite-size effects inherit a characteristic time dependence due to the dynamics of the cloud.

\subsubsection*{Spin-induced quadrupole} \label{sec:kappa}

A spinning compact object has a series of multipole moments~\cite{Geroch:1970cd, Hansen:1974zz}  (see Appendix~\ref{sec:eft}). For the Kerr BH, the no-hair theorem \cite{Carter:1971zc, Robinson:1975bv} implies that these moments are fixed completely by the mass and spin of the BH, while additional independent parameters are needed to characterize other objects, such as NSs~\cite{Laarakkers:1997hb,Pappas:2012ns} or boson stars \cite{Ryan:1996nk}.  Given the mass quadrupole moment $Q$, it is customary to introduce the dimensionless parameter
\beq
 \kappa \equiv - \frac{Q M}{J^2} \,, \label{eqn: spin-induced Q def}
\eeq  
where $M$ and $J$ are the mass and angular momentum of the body. For Kerr BHs, we have $\kappa = 1$~\cite{Hansen:1974zz}, while for NSs, $\kappa \approx 1.4 - 8 $, depending on the equation of state~\cite{Laarakkers:1997hb, Pappas:2012ns}. The observation of a compact object with deviations from $\kappa=1$, and masses larger than the estimated upper bound of NSs (around $3M_\odot$)~\cite{kalogera}, would be direct evidence for the existence of exotic objects in nature, see e.g.~\cite{Krishnendu:2017shb}.  
The value of $\kappa$ can be obtained from the GW signal of the binary, through its effect on the phase of the signal~\cite{Poisson:1997ha,Porto:2016pyg}.  
The effect arises at 2PN order, i.e.~$\kappa\hskip 1pt v^4 \hskip 1pt \chi^2$, where $v$ is the typical relative velocity of the binary and $\chi \equiv J/M^2$ is the dimensionless spin parameter. 

\vskip 4pt
As long as the mass of the boson cloud constitutes a sizable fraction of the initial BH mass, the metric of the (isolated) BH-cloud would depart from that of the Kerr background. These departures include spin-induced multipole moments, which are not uniquely determined by the mass and spin of the cloud. A rigorous computation of these quantities would require incorporating the backreaction on the spacetime geometry, which is beyond the scope of this work.\footnote{Exact solutions with (complex) scalar hair around spinning BHs have been studied numerically in the literature, see e.g. \cite{Herdeiro:2014goa}. In principle, these quasi-stationary spacetimes can also have $\kappa \gg 1$.} 
Instead, we will assume that the cloud dominates the contribution to the multipole moments of the BH-cloud, and estimate $\kappa$ by comparing the cloud's mass quadrupole, $Q_c$, to $J^2/M$, where $M$ and $J$ are the {\it total} mass and angular momentum of BH-cloud system. We choose to normalize $\kappa$ with $(M, J)$ of the BH-cloud, instead of $(M_c, J_c)$ of the cloud itself, because {\it i}\hskip 1pt) the quantities $(M, J)$ can be directly measured through GW observations, and {\it ii}\hskip 2pt) they are conserved throughout the evolution (up to small losses due to GW emission). The parameters $(M, J)$ therefore also coincide with the mass and angular momentum of the \textit{initial} BH prior to the formation of the cloud.\vskip 4pt 
From the stress-energy tensor of the scalar field, we find for the $\ket{211}$ state (see Appendix~\ref{sec: BH-Cloud System})
\beq
 \kappa(\alpha)\, \geq \, - \frac{Q_c(\alpha)}{M^3} \sim  10^3 \left(\frac{M_c(\alpha)/M}{0.1}\right)\left(\frac{0.1}{\alpha}\right)^4\, ,
\eeq 
where we have imposed the weak cosmic censorship condition, $J \leq M^2$ \cite{1969NCimR...1..252P}, to obtain the lower bound. 
The effect on the GW phase then scales as
\beq
\kappa(\alpha) \hskip 1pt v^4 \hskip 1pt \chi^2\, \gtrsim \, 10^{-2} \left(\frac{M_c(\alpha)/M}{0.1}\right) \left(\frac{v}{\alpha/2}\right)^4  \, ,\label{kappac}
\eeq 
where we used $\chi \simeq 1$, and assumed that the initial BH is rapidly rotating (which is required for the cloud to form in the first place). 
Notice that, in the regime of validity of the perturbative expansion, the relative velocity satisfies  $v \lesssim \alpha/2$.\footnote{For $q \lesssim 1$, the virial theorem implies $v \simeq (\alpha/2) \sqrt{r_c/R_*}$ and hence $v \lesssim \alpha/2$ in the regime of validity of the multipole expansion, $r_c < R_*$.}
 For $v > \alpha/2$, the companion experiences a smaller amount of the cloud according to Gauss' law. Even though larger relative velocities may seem favorable, the reduction in the effective mass of the cloud will dominate, leading to negligible finite-size effects once the companion enters the BH-cloud system. 
\vskip 4pt
The existence of resonances in the orbital dynamics can lead to an abrupt drop, or significant change, in the mass of the cloud $M_c$, with a corresponding frequency-dependent variation of the GW signal. As the binary scans through the Rabi frequency, we have 
\beq
\frac{\kappa(t)-1}{\kappa(t_0)-1} \sim \frac{M_c(t)}{M_c(t_0)} \sim e^{- 2\mathcal{A}(t, t_0)}\, .
\label{eq:timek}
\eeq
This time dependence constitutes a distinctive signature of the existence of a BH-cloud in a binary system and may leave a measurable imprint in the waveforms. We will return to this in~\S\ref{sec: Phenomenology}.

\subsubsection*{Tidal deformations}

In addition to the permanent multipole moments, a compact object may also acquire induced multipoles in the presence of a gravitational perturbation. It is an interesting (and somewhat puzzling) fact that for BHs the tidal response to an external field vanishes  in four-dimensional Einstein gravity.\footnote{So far, this has been shown to hold for non-rotating BHs. However, it is also expected to be the case for Kerr BHs. See e.g.~\cite{pani} and references therein for progress along this direction.} This is often parameterized in terms of the tidal Love numbers,\footnote{These Love numbers are the analog of susceptibilities in electrodynamics, which describe the response of a distribution of charge to an applied electric (or magnetic) field.} 
which are zero for black holes \cite{Porto:2016zng,Damour:2009vw,Binnington:2009bb,Kol:2011vg,Gurlebeck}, but may have sizable values for neutron stars \cite{Binnington:2009bb,Damour:2009vw} and other more exotic objects (like boson stars)~\cite{Mendes:2016vdr,Cardoso:2017cfl,Sennett:2017etc}.  While the Love numbers first enter in the GW phase at 5PN order~\cite{damour,nrgr,Flanagan:2007ix, Hinderer:2007mb}, the lack of  standard model background in Einstein's theory offers a powerful opportunity to probe the dynamics of vacuum spacetimes in General Relativity, through precision GW~data. See \cite{Porto:2016zng} for more on this issue. 

\vskip 4pt  
Since the cloud is much less compact than an isolated BH, but carries a significant fraction of the mass and angular momentum of the system, we can,  in principle, have large Love numbers for the BH-cloud system.  Although a detailed computation is beyond the scope of this work, on dimensional grounds we expect the `tidal deformability' parameter, $\Lambda$, to scale as (see Appendix~\ref{sec:eft}) 
\beq
\label{tildece}
\Lambda(\alpha)  \,\sim\, \left(\frac{M_c(\alpha)}{M}\right) \left(\frac{r_c}{2r_g}\right)^4 \,\sim\, 10^7 \left(\frac{M_c(\alpha)/M}{0.1}\right) \left(\frac{0.1}{\alpha}\right)^8  \, .
\eeq
The parameter $\Lambda$ enters in the phase of the waveform at 5PN order, and its imprint on the signal from the binary therefore scales as
\beq
\Lambda(\alpha) \hskip 2pt v^{10} \sim 10^{-6} \left(\frac{M_c(\alpha)/M}{0.1}\right) \left( \frac{\alpha}{0.1}\right)^2 \left( \frac{v}{\alpha/2} \right)^{10} \,,
\eeq
where $v \lesssim \alpha/2$ in the regime of validity of the multipole expansion. 
\vskip 4pt
As for the spin-induced quadrupole, the presence of the binary companion can induce 
a characteristic time dependence of the Love numbers 
\beq
\frac{\Lambda (t)}{\Lambda(t_0)} \sim \frac{M_c(t)}{M_c(t_0)} \sim  e^{-2 \mathcal{A}(t, t_0)}\,, 
\label{equ:LoveT}
\eeq
most notably for Bohr resonances, which occur at shorter binary separations. 

\vskip 4pt 
As the companion progresses into the cloud and approaches the merger with the central BH, both the effects of the spin-induced quadrupole and Love numbers will be dominated by the Kerr solution. In particular, the contribution from the Love number will become negligible. By comparing the values obtained before and after the resonance, 
 the variation of the spin-induced quadrupole and Love number(s) can turn into a key signature of the presence of a cloud in a binary system.

 \subsection{Probing Ultralight Scalars} 
 \label{sec: Phenomenology}

In this section, we study the new ways in which ultralight scalars can be probed with binary inspirals. We will concentrate on paradigmatic GW sources for which both the hyperfine and Bohr resonances can play a major role, accessible to present and planned GW detectors. 

\vskip 4pt
To be able to probe the resonances, the lifetime of the cloud has to be sufficiently long, and the resonance sharp enough, to be detected within the operational time window of GW observatories:
\begin{enumerate}

\item \textit{Cloud lifetime}. 
Survival of the cloud until the resonance typically requires $\tau_c \gtrsim 10\,{\rm Myr}$ for stellar-mass binaries, using generic lower bounds on their merger times~\cite{TheLIGOScientific:2016htt}. 
For supermassive BHs, we conservatively take $\tau_c \gtrsim 1$\,Gyr to account for the large uncertainties in the formation channels.
\item \textit{Resonant decay}.  To observe significant resonant decay in the GW signal by present and future GW detectors, the typical decay time, $\tau_d$, must be $\lesssim$ a few years.
\end{enumerate}
Using (\ref{equ:tauc}) and \eqref{eqn: resonance decay timescale} for $\tau_c$ and $\tau_d$, respectively, these considerations typically select narrow ranges of $\alpha$, depending on $M$.  
For stellar-mass BHs, we will be sensitive to $\alpha \simeq 0.07$ for real fields, while axisymmetric clouds made out of complex fields can be observed for larger values of $\alpha$. On the other hand, for supermassive BHs with $M \gtrsim 10^5 M_\odot$, the constraint on the decay time $\tau_d \lesssim$\,years 
 implies $\alpha \gtrsim 0.2$, with larger values of $\alpha$ required for larger $M$. Even though this typically leads to $\tau_c \lesssim 1$\,Myr for real fields, astrophysical processes such as accretion may alleviate the stability issue~\cite{Brito:2014wla}.\footnote{In general, we can relax the condition (\ref{eqn: ttm vs lifetime}) for the cloud's stability by requiring the lifetime to be sufficiently long to reach the resonance after formation, e.g.~$\tau_c \gtrsim 1$\,Myr, but not necessarily to last until merger.}

\vskip 4pt 
In the following, we will discuss how boson clouds around BHs can be probed in continuous-wave and binary searches. In the former, the resonant decay of the monochromatic signal is surveyed with ground-based detectors and LISA, while in the latter, the focus is on the resonant depletion/attenuation of finite-size effects, observed through precision measurements in the GW emission from the two-body system with LISA.

\subsubsection*{Continuous-wave searches}
 \label{sec: continuous wave search}  

 \begin{figure}[t!]
\centering
\includegraphics[scale=1, trim = 0 0 0 0]{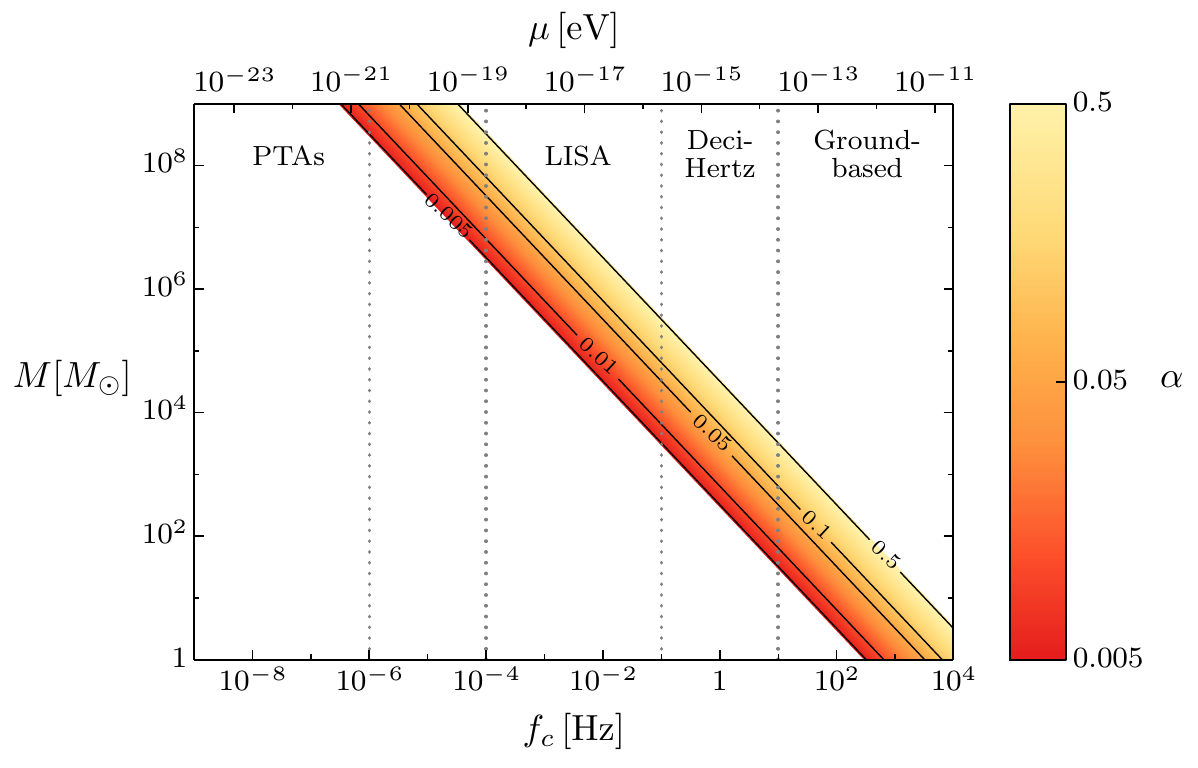}
\caption{Frequency of the monochromatic signal for  physically motivated ranges of $\alpha$ and $M$. The vertical dotted lines illustrate the typical observational bands of various GW observatories.} 
\label{fig: faPlot}
\end{figure}
 
The frequency of the signal from the cloud is shown in Fig.~\ref{fig: faPlot}, for physically motivated values of $\alpha$ and $M$, cf.~\eqref{eqn: GW annihilation frequency}, and compared to the frequency bands of ground-based~\cite{TheLIGOScientific:2014jea, TheVirgo:2014hva, Somiya:2011np, Sathyaprakash:2012jk, Evans:2016mbw} and space-based~\cite{AmaroSeoane:2012km, Kawamura:2011zz, Graham:2017pmn} detectors, as well as pulsar timing arrays (PTAs)~\cite{Hobbs:2013aka, 0264-9381-30-22-224009, 0264-9381-30-22-224008, 0264-9381-30-22-224010, Smits:2008cf, Hobbs:2014tqa}. We see that ground-based experiments probe  $M \lesssim 3 \times 10^3 M_\odot$, while LISA would give us access to $M \gtrsim 3 \times 10^3 M_\odot$.  
 For LIGO, 
the majority of resolvable events with significant SNR are produced by galactic sources, cf.~(\ref{eqn: SNR  annihilation}). The present binary event rates inferred from LIGO \cite{Abbott:2016nhf} indicate that the most promising sources in the galaxy are far from merging. Indeed, population synthesis codes~\cite{Nelemans:2001nr, Belczynski:2008nh, Lamberts:2018cge} suggest that there are presently of the order of $10^5$ to $10^6$ galactic BH-BH and BH-NS binaries of comparable masses with orbital frequencies distributed in the mHz region.   
Provided a fraction of these binary systems transit through a resonant epoch, a sharp decay in the GW signal from the cloud may be observed. This is particularly suitable for the hyperfine decay, for which the orbital resonances for stellar-mass BH binaries fall naturally in the mHz region; see \eqref{eqn: GW resonance frequency} below.
However, due to the finite lifetime of the clouds, detecting this effect in the galaxy requires the clouds to be formed relatively recently, 
while the majority of these binaries may have formed much earlier \cite{Lamberts:2018cge}. Despite these inherent uncertainties in the binary formation mechanisms, there are fortunately a large number of galactic binaries, each sampling a distribution of initial conditions. This constitutes an \textit{ensemble} which allows for a `scanning' of the orbital frequencies for the resonance, and the corresponding depletion of the monochromatic GW signal.

\vskip 4pt 
The scanning for both the hyperfine and Bohr resonances is more promising with LISA, which would also be sensitive to extra-galactic sources. In particular, a sharp decay may be observed for clouds surrounding supermassive BHs with a small companion, in what are known as extreme-mass-ratio inspirals (EMRIs). Similarly to the tuning required for the binary signals from EMRIs to fall into the LISA band \cite{Gair:2004iv}, a companion may excite orbital resonances, inducing a sharp depletion on the GW signal from a supermassive BH-cloud. 
The formation mechanism and population of intermediate-mass and supermassive BHs, as well as EMRIs, is less well-understood. However, since LISA will be able to probe sources of extra-galactic origin, the search volume increases considerably, allowing for the detection of GW signals at higher redshifts.

\vskip 4pt
As a concrete example, consider a real scalar field with $\mu \simeq 10^{-17}$eV, forming a cloud surrounding a supermassive BH with mass $M \simeq 5 \times 10^6\, M_\odot$. For simplicity, let us assume that the BH accretes matter, so that the issue of the lifetime of the cloud may be ignored. In this scenario we have $\alpha \simeq 0.37$, and a resonant decay time of about a few months. The continuous monochromatic signal from the cloud occurs around $f_c \simeq 5 \,$mHz, the most sensitive part of the LISA band. 
For binaries that transit through the hyperfine resonance, the complete termination of the signal occurs for $q \lesssim 10^2$. The decay can arise either from significant perturbative depletion in the early stages or, provided that the initial orbital frequency $\Omega_0$ lies close to $\epsilon_h$, the cloud may survive the perturbative decay and experience a sharp depletion within a few months.\footnote{One can show that for binary systems with initial condition $\Omega_0 < \epsilon_h$, Fig.~\ref{fig:co-rotating} is modified such that regions~{\rm II, III$_a$} and~{\rm III$_b$} turn red, but not region~{\rm III$_b$}, making almost the entire parameter space susceptible to decay. }  
On the other hand, significant attenuation of the signal due to Bohr resonances requires $q < 1$, cf.~Fig.~\ref{fig:counter-rotating}. These type of binary systems, scanning over the space of orbital resonances, are expected to be abundant in nature, which motivates comprehensive continuous-wave searches with space-based interferometers. 

\subsubsection*{Binary searches}

\begin{figure}[t]
\centering
\includegraphics[scale=1, trim = 0 0 0 0]{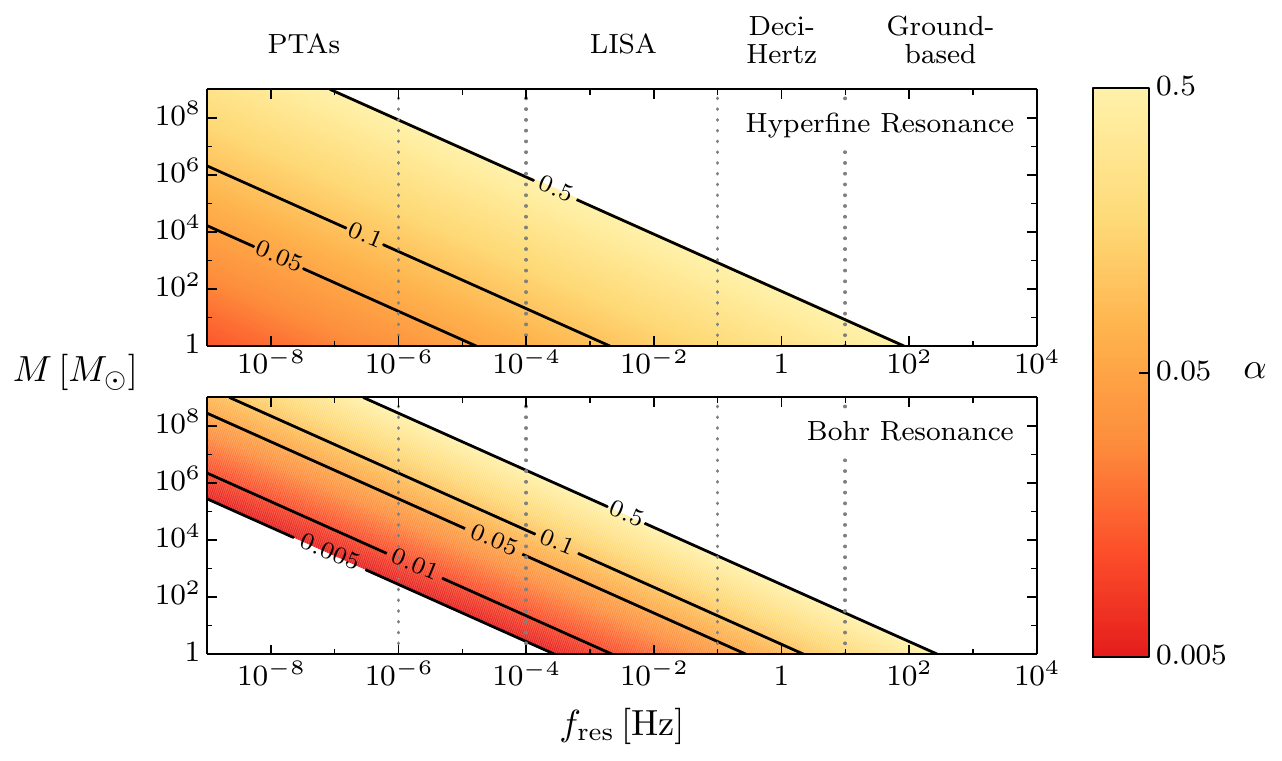}
\caption{The resonance frequency $f_{\rm res}$ for physically motivated ranges of $\alpha$ and $M$. The vertical dotted lines illustrate the typical observational bands of various GW observatories.  }
\label{fig: fResPlot}
\end{figure}

Precision finite-size measurements in binary searches are a new probe of boson clouds around~BHs. If the resonance falls within the frequency window of future observatories, it becomes a distinctive feature of the GW signal.  To see when that may be the case, it is instructive to translate the orbital frequency at which the resonance transitions occur, $\Omega_{\rm res}$, to the GW frequency, $f_{\rm res}$, emitted from the binary during the inspiral: 
\beq
f^{(i)}_{\rm res} \equiv \frac{\Omega^{(i)}_{\rm res}}{\pi} = \frac{|\epsilon_i|}{\pi} \, \simeq\,
\begin{dcases}
\ 7.2\, \text{mHz} \ \frac{1}{1+4 \alpha^2} \left(\frac{\alpha}{0.1} \right)^7 \left(\frac{3 M_\odot}{M} \right)  , & \quad \text{Hyperfine} \\[4pt]
\ 0.75 \,  \text{Hz} \ \left(\frac{\alpha}{0.1} \right)^3 \left(\frac{3 M_\odot}{M} \right) . & \quad \text{Bohr} \label{eqn: GW resonance frequency}
\end{dcases}
\eeq 
This frequency is shown in Fig.~\ref{fig: fResPlot}, for physically motivated ranges of $\alpha$ and $M$. We see that, for most cases, both the hyperfine and Bohr frequencies are too low to be observed by ground-based detectors.  The best prospects for detecting the resonance feature in future binary searches is therefore through space-based detectors, such as LISA.

\vskip 4pt
The targeted values of $\mu$ (and $\alpha$) for real scalars is restricted mostly by the stability of the cloud. 
For axisymmetric configurations with complex fields, GW emission from the cloud is not present and therefore binary searches are perhaps the only way in which these bosons could be potentially observable.\footnote{It is worth mentioning that in previous proposals, probes of scalar fields in binary systems rely on the formation of boson stars~\cite{Sennett:2017etc}. However, unlike the BH-cloud, 
boson stars do not experience resonant depletion, which therefore is a unique of feature of boson condensates surrounding spinning BHs.} As we see in Fig.~\ref{fig: fResPlot}, observing the resonant decay is most promising in the deci-Hertz and LISA frequency bands. For simplicity, let us focus on sources that are observable by LISA, which is sensitive to compact binaries with large masses $M \in [10^4, 10^7]\, M_\odot$, although in principle it can also detect stellar-mass BHs~\cite{Sesana:2016ljz}. Binary searches can then broadly be divided into two classes: {\it i}\hskip 1pt) binaries of comparable masses, $q \in [10^{-2}, 10^2]$, and {\it ii}\hskip 1pt)~EMRIs, with $q \in [10^{-6}, 10^{-2}] \cup [10^{2}, 10^{6}]$.

\vskip 4pt In the case of comparable-mass binaries, 
finite-size effects manifest themselves at high PN orders, and therefore become accessible when the relative velocity of the binary becomes sizable. Once the companion enters the cloud, the finite-size effects will be dominated by the BH solution, for instance the spin-induced quadrupole will be given by $\kappa \simeq 1$ and the Love numbers will vanish. To detect the time-dependent effects in \eqref{eq:timek} and \eqref{equ:LoveT}, the imprints in the GW phase have to be observable near the resonance. 
For co-rotating orbits, where the hyperfine resonance occurs at larger distances, measuring the depletion of the finite-size effects is challenging. Nevertheless, it is also possible that the cloud forms in a binary systems with $\Omega_0 > \epsilon_h$, therefore missing the hyperfine resonance. While level mixings would still deplete the cloud, it is possible that it survives until later times; e.g.~see Fig.~\ref{fig:co-rotating} for $\alpha \simeq 0.1$ and $q \simeq 10^{-2}$. In this scenario, resonant depletion would not occur and finite-size effects may still be observed at later stages of the inspiral.

\vskip 4pt
Detecting the resonant decay is more promising for counter-rotating orbits, where the resonance generically occurs at shorter separations. In that case, the relative velocity near the resonance is given by $v_{(b)} \simeq 0.65\,(\alpha/2)$, which can be significant. For example, for $\alpha \simeq0.1$ and $M_c \simeq 0.1 M$, the finite-size effects near the Bohr resonance are of order, 
\begin{align}
\kappa(\alpha) \hskip 1pt v_{(b)}^4 \hskip 1pt \chi^2 &\,\sim\,  10^{-3}\,,  \\
\Lambda(\alpha) \hskip 1pt v_{(b)}^{10} &\,\sim\, 10^{-8} \, . 
\end{align}
The resonant depletion of these quantities can be as large as $\simeq 20 - 30\%$ (see Fig.~\ref{fig:counter-rotating} for $\alpha \simeq 0.1$ and $q \simeq 10^{-2}$).  
Notice that, while small, both types of finite-size effects are larger than the usual 2PN and 5PN terms, respectively. In particular, 
near the resonance, tidal effects are comparable to typical 3PN contributions and the corrections to the GW phase appear to be within reach of future GW observatories such as LISA~\cite{Cardoso:2017cfl}.

\vskip 4pt
In the case of EMRIs, the evolution of the companion in the background of a supermassive BH can be obtained through perturbative self-force computations \cite{Barack:2009ux,Galley:2013eba}. The tidal deformability of the object is typically not taken into account, since it is highly suppressed when $M_* \ll M$. Nonetheless, systems with $q \ll 1$ still offer an excellent probe of deviations from $\kappa = 1$ for the background spacetime. It has been estimated that a variation in the spin-induced quadrupole of order $\Delta \kappa/\kappa \simeq 10^{-2}$ may be detectable by LISA~\cite{Barack:2006pq}. For example, counter-rotating binaries with $\alpha \simeq 0.1-0.4$ and $q \lesssim 10^{-2}$ would behave in a similar way as the blue curves in Fig.~\ref{fig:counter-rotating}. 
This strongly motivates searches for large spin-induced quadrupole moments in EMRIs and their resonant depletion.

\section{Conclusions and Outlook} 
\label{sec:Conclusions}

Astrophysical BHs exist with a wide range of masses, from a few to billions of solar masses.  
When these BHs are rapidly rotating, they can produce condensates of ultralight bosons with masses in the range $[10^{-20}, 10^{-10}] \,{\rm eV}$, which includes well-motivated candidates for physics beyond the Standard Model, such as the QCD axion~\cite{PhysRevLett.38.1440, PhysRevLett.40.223, PhysRevLett.40.279}, axion-like particles in string theory~\cite{Svrcek:2006yi, Baumann:2014nda, Arvanitaki:2009fg}, and interesting new possibilities for dark matter~\cite{Hui:2016ltb}. Extremely ambitious observational programs are under way to search for these light particles in the lab~\cite{Essig:2013lka}, in astrophysics~\cite{Raffelt:1996wa} and in cosmology~\cite{Abazajian:2016yjj, Baumann:2016wac, Brust:2013xpv}.  As impressive as these efforts are, they are blind to particles that couple to ordinary matter only with gravitational strength. Such extremely weakly coupled particles would nevertheless be produced by BH superradiance. The resulting boson clouds can be long-lived (on cosmological/astrophysical timescales) creating temporary BH hair that can be searched for in future GW observations, either in isolation or as a member of a binary system.

\vskip 4pt
In this paper, we studied the dynamics of scalar condensates in binary inspirals, both with real and complex fields. We showed that the presence of a companion greatly enriches the dynamics of the system, by inducing mixing between growing and decaying modes of the cloud. At certain critical orbital frequencies the evolution becomes non-perturbative, leading to enhanced mixing through resonances with the orbital motion of the system. By restricting to simplified two- and three-state models of the cloud, we have shown that the resulting time dependence can significantly affect physical observables through a sharp depletion (or at least an attenuation) of the energy density in the bosonic field. This has important phenomenological consequences, both for the monochromatic signal emitted by the cloud and the finite-size effects imprinted in the waveforms of the binary signal. The characteristic time dependence of the signals thus become a very distinctive feature of the dynamics of boson clouds in binary systems, which would not be present in other scenarios, e.g.~boson stars.

\vskip 4pt 
The results presented here should be considered a first step towards more accurate descriptions of the dynamics of the cloud. For instance, a proper calculation of the finite-size effects associated with the BH-cloud will require numerical simulations, to incorporate the backreaction on the background geometry. We have also ignored the backreaction on the BH mass and spin due to the evolution of the surrounding cloud. Moreover, our analytical treatment breaks down when higher modes are excited during the resonances, as well as during the late stages of the inspiral, where the dynamical system becomes strongly time dependent. In these cases, only numerical approaches are able to properly inform us about the exact details of the dynamics. Other aspects of our estimations can also be improved upon. For example, we have restricted our analysis to quasi-circular orbits on the equator. This was only for simplicity, and it would be relatively straightforward to extend our analysis to the case of elliptic orbits. Finally, we did not perform a detailed forecast of the expected event rates, and signal strengths, with future GW observatories. This requires a careful treatment of formation mechanisms, as well as the associated  generation of boson clouds. 

\vskip 4pt 
In spite of our simplifications, we have identified robust GW signatures for clouds in binary inspirals, which open up new ways of probing physics beyond the Standard Model through current and future GW observations. In particular, we hope that future measurements could not only detect these ultralight particles, but also help us elucidate their putative properties. Following an analogy with collider physics, we expect to extract not only the particle's masses, but also their spins and self-couplings~\cite{VectorSuper}, ultimately realizing the idea of the `gravitational collider'. 

	\vskip23pt
	\subsection*{Acknowledgements}
	We thank Mina Arvanitaki, Masha Baryakhtar, Alessandra Buonanno, Sam Dolan, William East, Lotte ter Haar, Tanja Hinderer, Badri Krishnan, Luis Lehner, Ilya Mandel, Mehrdad Mirbabayi, David Nichols, Samaya Nissanke, Paolo Pani, Frans Pretorius, John Stout, Ka Wa Tsang, Benjamin Wallisch, and Matias Zaldarriaga for helpful comments and discussions. We thank Benjamin Wallisch for assistance in producing some of the figures. D.\,B.~is grateful to the Leung Center for Cosmology and Particle Astrophysics at the National Taiwan University and the Institute for Advanced Study for hospitality while some of this work was being performed.  The work of D.\,B.~is supported by a Vidi grant of the Netherlands Organisation for Scientific Research~(NWO) that is funded by the Dutch Ministry of Education, Culture and Science~(OCW). The work of H.\,S.\,C.~is supported by NWO. The work of R.A.P. was supported by the Simons Foundation and S\~ao Paulo~Research Foundation Young Investigator Awards, grants~2014/25212-3 and 2014/10748-5, and by the German Science Foundation (DFG) within the Collaborative Research Center (SFB) 676 `Particles, Strings and the Early Universe.' R.A.P. would like to thank the International Center for Theoretical Physics (ICTP) in Trieste for hospitality while this work was being completed. R.A.P. is grateful to Leonidas K. Porto, Emiliano A. Porto and Imme Roewer for the unconditional support. 
	
	\clearpage
	\appendix

\section{Gravitational Atom}
\label{sec: BH-Cloud System}

In this appendix, we derive some of the properties of the BH-cloud presented in the main text. In particular, we compute the corrections to the frequency eigenvalues of the Hamiltonian in the non-relativistic limit. We also determine the mass quadrupole moment of the BH-cloud in the Newtonian approximation.
Unless stated otherwise, these results apply to both real and complex scalar fields.

\subsubsection*{Kerr geometry} 

The Kerr metric in Boyer-Lindquist (BL) coordinates is
\beq
\d s^2 = - \frac{\Delta}{\rho^2}\left(\d t - a \sin^2 \theta\, \d \phi \right)^2 + \frac{\rho^2}{\Delta} \d r^2 + \rho^2\, \d \theta^2 + \frac{\sin^2 \theta}{\rho^2}  \left(a \d t - (r^2 + a^2) \, \d \phi \right)^2\, , \label{equ:Kerr}
\eeq
where we have defined
\beq
\Delta \equiv r^2 -  2 Mr +a^2  \quad {\rm and} \quad \rho^2 \equiv r^2 + a^2 \cos^2 \theta\, .
\eeq
Since the metric components have no explicit dependence on $t$ and $\phi$, the Kerr spacetime is stationary and axisymmetric.
The roots of $\Delta$ determine the event horizon at $r_+ = M + \sqrt{M^2 -a^2}$ and the Cauchy horizon at $r_-= M-\sqrt{M^2-a^2}$. The surface $g_{tt}=0$ at $r_{\rm E} = M +\sqrt{M^2-a^2 \cos^2 \theta}$ is the ergosphere.
The angular velocity of the spinning BH at its event horizon is
\beq
\Omega_{\rm H} = \frac{a}{2 M r_+}\, .
\eeq
The BH angular velocity $\Omega_{\rm H }$ plays a crucial role in the superradiance condition (\ref{eqn: superradiance condition}).

\subsubsection*{Test scalar field}

The quadratic Lagrangian of a real scalar field is\footnote{Since a complex scalar field may be described in terms of two real fields, we will restrict the following discussion to the real case only.}
\beq
\mathcal{L} \,=\,  - \frac{1}{2} g^{ab} \nabla_a \Psi \nabla_b \Psi - \frac{1}{2} \mu^2 \Psi^2\, . 
 \label{eqn:free scalar field action}
\eeq
Ignoring the backreaction of the field's stress-energy on the metric, we can use the Kerr solution~(\ref{equ:Kerr}) for the metric $g^{ab}$. Expressed in BL coordinates, the Klein-Gordon equation is separable into a set of ordinary differential equations.  To find the stationary solutions of the KG equation, we consider the following ansatz
\beq
\Psi(t,r,\theta,\phi) \,=\,   \sum_{\ell m} \int \frac{\d \omega}{2\pi} \ \text{Re} \left[ e^{-i \omega t} e^{i m \phi} R_{\omega \ell m}(r) S_{\ell m}(\theta, c)  \right]  .   \label{eqn: Psi decomposition}
\eeq
The plane wave solutions in $t$ and $\phi$ are a direct consequence of the isometries of the Kerr solution. The angular functions $S_{\ell m}(\theta, c)$ are called spheroidal harmonics with {spheroidicity parameter} $c^2 \equiv a^2(\omega^2-\mu^2)$. In the limit $c^2 \to 0$, we have $S_{\ell m}(\theta) e^{i m \phi} \to Y_{\ell m}(\theta, \phi)$, where $Y_{\ell m}$ are the ordinary spherical harmonics. The radial functions $R_{\omega  \ell m}(r)$ do not admit analytic solutions and have to be determined numerically. However, as we shall see, at leading order in an expansion in $r^{-1}$, they take the form of the radial functions of the hydrogen atom. 

\subsubsection*{Non-relativistic limit}

The ansatz (\ref{eqn: Psi decomposition}) corresponds to stationary solutions of a massive scalar field around the BH. To study the \textit{dynamical} evolution of $\Psi$, it proves useful to make  another ansatz
\beq
\Psi (t, \textbf{r})=  \frac{1}{\sqrt{2\mu}} \left[ \psi (t, \textbf{r})\, e^{-i \mu t} + \psi^* (t, \textbf{r}) \, e^{+ i \mu t}  \right]  , \label{eqn: real scalar field 2}
\eeq
where $\psi$ is a complex scalar field which varies on a timescale that is much longer than $\mu^{-1}$. The action for $\psi$ reads
\begin{align}
S = \int \d^4 x \sqrt{-g} \left( - \frac{1}{2 \mu} \left[  \nabla_a \psi^* \nabla^a \psi + i \mu g^{0 a}  \left( \psi^* \nabla_a \psi -  \psi \nabla_a \psi^* \right) + \mu^2 ( g^{00} + 1) \psi^* \psi \right]  \right)  , \label{eqn: psi action}
\end{align}
which spontaneously breaks time reparametrization. 
Expanding in powers of $r^{-1}$, we obtain the following effective action 
\beq
S  = \int \d t \d r \d\theta \d\phi \, r^2 \sin \theta\, \Big[ \mathcal{L}_2 + \mathcal{L}_4 + \mathcal{L}_5 + \cdots  \Big]\,  , \label{eqn: action large distance}
\eeq
where $\mathcal{L}_n$ denotes terms of order 
$\mathcal{O}(\alpha^n)$, 
\beq
\begin{aligned}
\mathcal{L}_2 & \equiv i \psi^* \partial_t \psi - \frac{1}{2 \mu }  \partial_r \psi^* \partial_r \psi - \frac{1}{2\mu r^2 }  \partial_\theta \psi^*  \partial_\theta \psi -  \frac{1 }{2\mu r^2 \sin^2 \theta }\partial_\phi \psi^* \partial_\phi \psi + \frac{\alpha}{r}  \psi^* \psi    \, ,  \\
\mathcal{L}_4 & \equiv  \frac{1}{2\mu} \partial_t \psi^* \partial_t \psi +  \frac{2 M }{r } \left[ i  \psi^* \partial_t \psi  + \frac{1}{ 2 \mu }  \partial_r \psi^* \partial_r \psi   + \frac{ \alpha}{r} \psi^* \psi  \right]  ,   \\
\mathcal{L}_5 & \equiv \frac{2 i a M  \psi^* \partial_\phi \psi }{r^3 } \, . \label{eqn: Kerr-scalar large r lagrangian}
\end{aligned}
\eeq
For the power counting, we have used the fact that $\alpha < 1$ and $r \sim r_c \simeq (\mu \alpha)^{-1}$. 
The Lagrangian $\mathcal{L}_2$  defines the non-relativistic limit, which leads to the Schr\"odinger equation
\begin{align}
i \frac{\partial }{\partial t} \psi(t, \textbf{r}) = \left[ - \frac{1}{2\mu} \nabla^2 - \frac{\alpha}{r} \right] \psi(t, \textbf{r}) \, .  \label{eqn: Schrodinger like with gravity}
\end{align}
This is analogous to the non-relativistic limit of the hydrogen atom. By expanding $\psi$ in terms of the stationary eigenstates 
\begin{align}
\psi(t, \textbf{r}) = \int \frac{\d\omega}{2\pi}  \sum_{\ell m}  \psi_{n \ell m}(t, \textbf{r}) \, ,
\end{align}
and using~(\ref{eqn: Psi decomposition}), we can identify 
\begin{align}
\psi_{n \ell m} (t, r, \theta, \phi)  \simeq  e^{-i (\omega - \mu)t}  \bar{R}_{n \ell}(r) Y_{\ell m}( \theta, \phi)  \, , \label{eqn: Bohr radial eigenfunction}
\end{align}
where we have used the small spheroidicity approximation $S_{\ell m}(\theta) e^{i m \phi} \simeq Y_{\ell m}(\theta, \phi) + \mathcal{O}(\alpha^2)$, and defined $\bar{R}_{n \ell}(r) \equiv \sqrt{\mu/2}  R_{\omega  \ell m}(r)$. Since we have expanded the action in (\ref{eqn: action large distance}) at large distances, the boundary condition at the BH event horizon is replaced by a regular boundary condition at the origin. The radial function $\bar{R}_{n \ell}(r)$ therefore take the form of the  radial functions of the hydrogen atom. 

\vskip 4pt
The Lagrangians $\mathcal{L}_4$ and $\mathcal{L}_5$ are higher-order self-interaction terms. Using the non-relativistic equation of motion~(\ref{eqn: Schrodinger like with gravity}), we can substitute $i \partial_t \psi$ and $\partial_\phi \psi = i m \psi$ back into $\mathcal{L}_4$ and $\mathcal{L}_5$. This gives
\begin{align}
\mathcal{L}_4  + \mathcal{L}_5 = \frac{1}{2\mu} \partial_t \psi^* \partial_t \psi +  \frac{\alpha }{\mu^2 r^3} \psi^* \big[  \ell(\ell +1)   - 2 r \partial_r  (r \partial_r)  - 2 \tilde{a} m \alpha  \big]    \psi  \, ,
\end{align}
where we have used the eigenvalue equation of the spherical hamornics
\beq
\left( \frac{1}{\sin\theta} \frac{\partial }{\partial \theta}  \sin \theta \frac{\partial}{\partial \theta} + \frac{1}{\sin^2 \theta}\frac{\partial^2}{\partial^2 \phi} \right) Y_{\ell m}(\theta, \phi) = - \ell (\ell + 1) Y_{\ell m} (\theta, \phi) \, .
\eeq
Finally, it is convenient to write the dynamics of $\psi$ in terms of the 
 Hamiltonian
\begin{align}
H_c  & \equiv T_c + V_c 
  \, ,\label{eqn: Hs}
\end{align}
where we have separated $H_c$ into its kinetic and potential components,
\beq
\begin{aligned}
T_c & \equiv  - \frac{1}{2\mu} \nabla^2 + \frac{1}{2\mu}  \partial^2_t  \, , \\
V_c & \equiv  - \frac{\alpha}{r} -  \frac{\alpha }{\mu^2 r^3} \left[\l(\l +1)   - 2 r \partial_r  (r \partial_r) \right]    +  \frac{2 \tilde{a} m \alpha^2 }{\mu^2 r^3}  \, . \label{eqn: TsVs}
\end{aligned}
\eeq
We see that the eigenfunctions of the cloud are determined by the Hamiltonian of a test particle of mass $\mu$. Furthermore, since the Kerr metric is stationary, the kinetic energy of the BH vanishes. However, as we will see in Appendix~\ref{sec: Gravitational Perturbation Appendix}, this contribution becomes important when the BH-cloud is part of a binary system.   

\subsubsection*{Frequency eigenvalues}
 
Including the mass term $\mu$ that was factored out in (\ref{eqn: real scalar field 2}), the eigenfrequency spectrum of the Hamiltonian (\ref{eqn: Hs})  is
\beq
\omega_{n \ell m} = \mu \left(1 - \frac{ \alpha^2}{2 n^2} - \frac{\alpha^4}{8 n^4} + \frac{ \left( 2\ell - 3 n + 1 \right) \alpha^4 }{n^4  (\ell + 1/2) }   + \frac{2 \tilde{a} m  \alpha^5}{n^3 \l (\l + 1/2) (\l + 1)}\right) , \label{eqn: eigenfrequency 1}
\eeq
where we have used the following identities
\beq
\begin{aligned}
\left \langle r^{-1} \right \rangle_{n \ell} & = \frac{\mu \alpha}{n^2} \, ,  \\
\left\langle r^{-2} \right\rangle_{n \ell} & = \frac{\mu^2 \alpha^2 }{n^3 (\ell + 1/2)} \, , \label{eqn: expectation value identities} \\
\left\langle r^{-3} \right\rangle_{n \ell} & = \frac{\mu^3 \alpha^3 }{n^3 \ell (\ell + 1/2)(\ell + 1)} \, , \quad \ell > 0 \\
\left\langle r^{-2} \partial_r \left( r \partial_r  \right) \right\rangle_{n \ell} & = \frac{\mu^3 \alpha^3}{n^4} \left( \frac{\ell - n + 1/2}{\ell + 1/2} \right)  ,
\end{aligned} 
\eeq
with $\langle \cdots \rangle_{n \ell}$ defined as the expectation value with respect to~$\bar{R}_{n \ell}(r)$. 
Equation (\ref{eqn: eigenfrequency 1}) shows that the frequency eigenvalues of modes with different quantum numbers $\{ n \ell m \}$ are different. It is precisely these differences in the eigenfrequencies that allow level mixings to occur.

\subsubsection*{Klein-Gordon norm} \label{sec: KG Norm}

The action (\ref{eqn: psi action}) is invariant under the global $U(1)$ transformation $\psi  \to \psi \, e^{- i \beta}$, 
where $\beta$ is an arbitrary constant. The associated Noether current is 
\beq
J^a  = \frac{i}{2\mu} \left[ \psi \nabla^a \psi^* - \psi^* \nabla^a \psi \right] - g^{0a} \psi^* \psi \, .
\eeq
At leading order in the expansion in $r^{-1}$, the conserved charge is $J^0 \simeq \psi^* \psi$, which can be interpreted as the occupation density of a particular state. This is analogous to the probability density in quantum mechanics, after suitable normalization.

\subsubsection*{Quadrupole moment}

The conserved energy associated with the energy-momentum tensor of the scalar field is 
\begin{align}
E \equiv -  \int_{\Sigma} {T^0}_a k^a \, ,
\end{align}
where $k^a = {\delta^a}_0$ is the time-like Killing vector field of the Kerr metric and $\Sigma$ is a spacelike hypersurface of constant $t$. 
At leading order in $r^{-1}$ and $\alpha$, the energy density of the $\ket{211}$ mode of the cloud is 
\beq
\rho_E (t, \textbf{r}) \equiv - {T^0}_0  \simeq \frac{A^2 \mu^6 \alpha^5  }{64 \pi } \,  r^2 e^{ - \alpha  \mu r} \sin^2 \theta \, , \label{eqn: cloud energy density}
\eeq
where $A$ is the amplitude of the scalar field, which is determined by the efficiency of the superradiant instability.
At leading order in $\alpha$, the energy density $\rho_E$ is equal to the mass density $\rho_m$ of the cloud 
and the mass quadrupole moment can be approximated by
\begin{align}
Q_c = \int_0^\infty \d r \int_0^\pi  \d\theta \int_0^{2\pi} \d\phi \ r^2 \sin \theta  \, \left[  \rho_m r^2 P_2 (\cos \theta) \right]  , \label{eqn: cloud quadrupole moment}
\end{align}
where $P_2(\cos \theta)$ is the Legendre polynomial. Normalizing the amplitude $A$ in (\ref{eqn: cloud quadrupole moment}) by the mass of the cloud 
\beq
M_c = \int_0^\infty \d r \int_0^\pi  \d\theta \int_0^{2\pi} \d\phi \ r^2 \sin \theta   \, \rho_m\,  ,
\eeq
we obtain
\begin{align}
 Q_c \simeq - \frac{6 M_c}{\alpha^2 \mu^2} \simeq - \frac{3}{8} M_c r^2_c  \, . \label{eqn: Qc}
\end{align}
The negative sign arises as a direct consequence of the spinning motion of the cloud.

\newpage
\section{Free-Falling Clouds } 
\label{sec: Gravitational Perturbation Appendix}

In \S\ref{sec: gravitational perturbation}, we derived the effect of the companion on the BH-cloud by considering the perturbed metric in Fermi coordinates. In this appendix, we present an alternative perspective of this derivation by considering a \textit{three-body analogy}. Our main goal is to show how a fictitous dipole in the gravitational potential, generated by a change of coordinates, cancels out. 

\subsubsection*{Three-body analogy}

The eigenfunctions of the cloud are determined by the Hamiltonian $H_c$ of a test particle of mass~$\mu$, cf.~(\ref{eqn: Hs}). 
This is analogous to the Hamiltonian of the electron in a hydrogen atom: while the electron wavefunction, as a solution of the Schr\"odinger equation, has a wave-like distribution around its nucleus, it is treated as a point particle at the level of the Hamiltonian.  In the rest of this appendix, we will adopt this particle picture of the Hamiltonian. 
\vskip 10pt
\begin{figure}[thb]
\begin{center}
\includegraphics[scale=0.6]{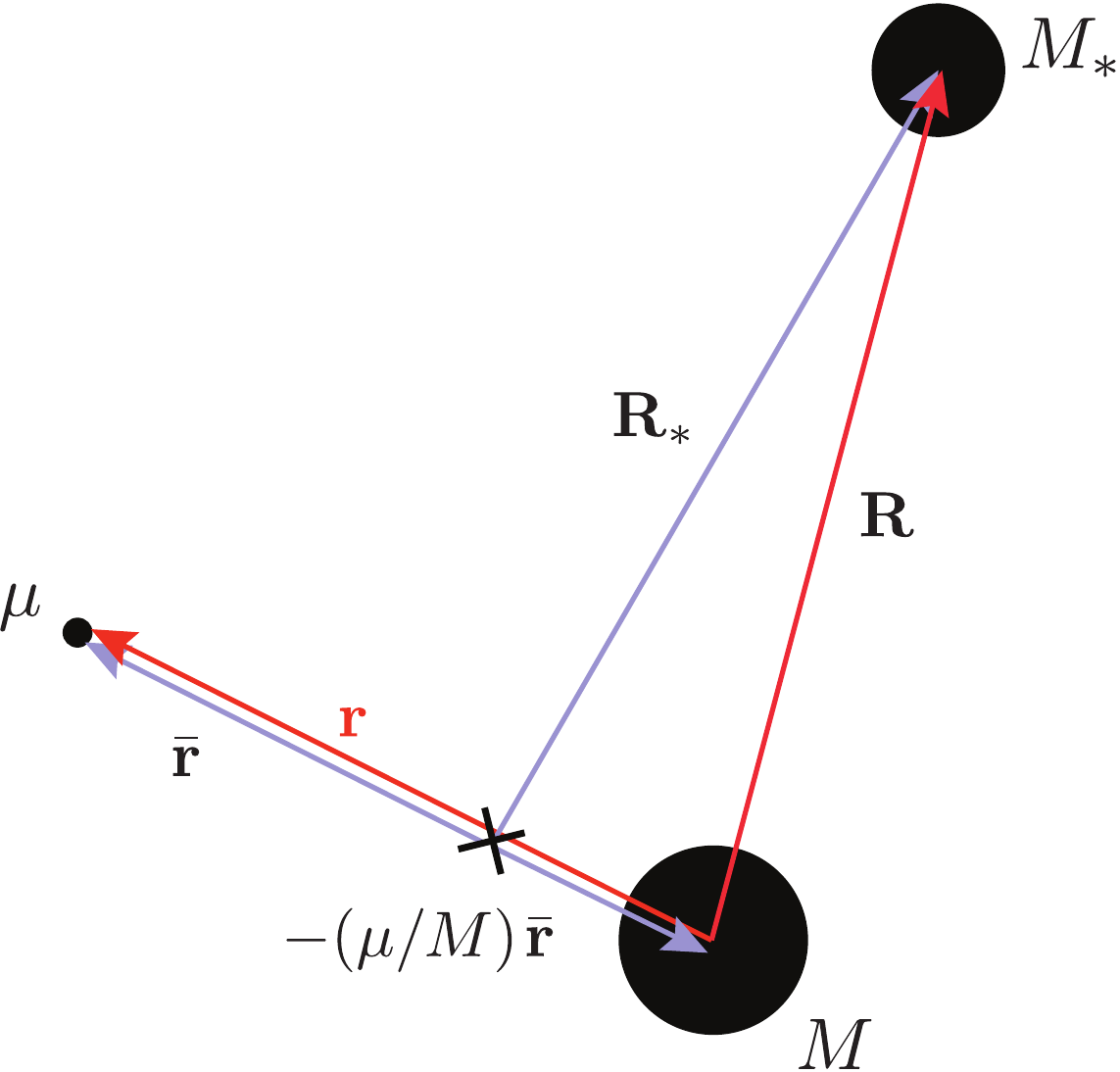}
\caption{Coordinates of the effective three-body system. } 
\label{fig:coordinates}
\end{center}
\end{figure}
\vskip -15pt
In the presence of a companion, of mass $M_*$, the BH-cloud will accelerate due to the external gravitational field. The Hamiltonian (\ref{eqn: Hs}) must hence be modified to include both the kinetic term of $M$ as well as the contributions from $M_*$. It is convenient to introduce the center-of-mass coordinate, $\boldsymbol{\rho} \equiv (M \r_1+\mu\r_2)/(M+\mu)$,
so that the total Hamiltonian $H_{\rm tot}$ of the three-body system can be written as 
\beq
H_{\rm tot} 
= \bigg[ \frac{\p_\rho^2}{2 (M + \mu)} + \frac{\p_r^2}{2\hat{\mu}}  + V_c(|\r|) \bigg] + \bigg[   \frac{\p_*^2}{2M_*} + V_*(\bar \r, \R_*) \bigg]\, ,
\eeq
where  $\r \equiv \r_2 - \r_1$ is the relative spatial separation between $\mu$ and $M$, and we introduced the reduced mass $\hat{\mu} \equiv M\mu/(M+\mu)$, the momenta  $\p_\rho \equiv (M+\mu) \dot{\boldsymbol{\rho}}$ and $\p_r \equiv \hat{\mu} \dot{\r}$, and the coordinates ${\bf R_*}\equiv \{R_*,\Theta_*,\Phi_*\}$ and $\bar \r \equiv \{\bar r, \bar \theta, \bar \phi\}$ relative to the center-of-mass 
(cf. Fig.~\ref{fig:coordinates}). 
Working in the Newtonian limit, the external potential $V_*$ is given by 
\beq
\begin{aligned}
V_*(\bar \r, \R_*) & = - \frac{M_* M}{|\R_* + \mu \bar \r/ M  |} - \frac{M_* \mu}{|\R_* -  \bar \r  |}  \label{eqn: Newtonian perturbation new 1}  \\
& = - M_*  \sum_{\ell_* m_*} \frac{4\pi}{2\ell_* + 1} \left( \frac{M (-\mu/M)^{\ell_* }+ \mu}{R_*} \right) \left( \frac{\bar r}{R_*} \right)^{\ell_*} \,Y^*_{\ell_* m_*}(\Theta_*, \Phi_*) \,Y_{\ell_* m_*}(\bar \theta, \bar \phi)   \, .  
\end{aligned}
\eeq
Substituting $\bar \r = (1+\mu/M)^{-1}\, \r$, and expanding in $\mu/M\ll 1$, we find 
\begin{align}
 V_*(r, R_*) &= - \frac{M_* (M+\mu) }{R_*} - \frac{M_* \mu}{R_*}\,  \sum_{\ell_* \ge 2} \sum_{|m_*| \le \ell_*} \frac{4\pi}{2\ell_*+1} \left(\frac{r}{R_*}\right)^{\ell_*}\, Y^*_{\ell_* m_*}(\Theta_*,\Phi_*) \,Y_{\ell_* m_*}(\theta,\phi)\, , \nonumber\\[4pt]
&\equiv  V_{*,0}(R_*) + \sum_{\ell_* \ge 2} V_{*, \ell_*}(r,R_*)\, .   \label{equ:dV2} 
\end{align}
The leading monopole term $V_{*, 0}$ determines the acceleration of the center-of-mass of the BH-cloud system.  It doesn't depend on the relative separation $\r$, so it doesn't lead to a shift in the energy levels or the mode functions of the system.  The remaining terms are a sum over harmonics, starting with the quadrupole $\ell_* =2$. Importantly, the dipole contribution $\ell_* =1$ vanishes in the center-of-mass frame.

\subsubsection*{Fictitious dipole}

We have shown that the contribution from the dipole vanishes in the centre-of-mass frame. By virtue of the equivalence principle, this property must hold for all coordinate systems. More generally, this is a manifestation of the fact that a constant gravitational gradient --- in this case the dipole term induced by the external gravitational field produced by the companion~$M_*$ --- is physically unobservable. We now show explicitly that this is indeed the case for the coordinate system centered at $M$.

\vskip 4pt
Consider expressing $H_{\rm tot} $ in terms of $(\r_1, \r)$ instead of $(\boldsymbol{\rho}, \r)$,
\beq
H_{\rm tot} = \left[  \left( 1 + \frac{\mu}{M} \right) \left( \frac{\textbf{p}_1^2}{2 M} + \frac{\textbf{p}_r^2}{2 \hat{\mu}}  \right)  + \frac{\mu}{ M \hat{\mu}} \textbf{p}_1 \cdot \textbf{p}_r  +   V_c(|\r|)   \right]  + \left[ \frac{\p_*^2}{2M_*} + V_*(\r, \R) \right]  ,
\eeq
where the gravitational potential in this choice of coordinates reads (cf.~Fig.\ref{fig:coordinates})
\beq
V_*(\r, \R)  = - \frac{M_* M}{|\R|} - \frac{M_* \mu}{|\R -  \r  |}   \, .
\eeq
Expanding $V_*$ for $r \ll R$ produces a dipole term
\beq
- \frac{M_* \mu}{|\R- \textbf{r}|} \supset - \left( \frac{M_* \mu}{R}  \right) \left( \frac{\hat{\textbf{n}} \cdot \textbf{r}}{R} \right)  , \label{eqn: gravitational dipole}
\eeq
where $\hat{\textbf{n}}$ is the unit vector along $\textbf{R}$, and we have used $|\textbf{R} - \textbf{r} | = R - \hat{\textbf{n}} \cdot \textbf{r} + \mathcal{O}(r^2)$. 
We will now show that this dipole is cancelled by the kinetic mixing term $ \mu \, \textbf{p}_1 \cdot \textbf{p}_r / M \hat{\mu} $. This is manifested most transparently in the $M_* / R \gg V_*$ limit, in which the gravitational attraction between $M$ and $\mu$ is negligible, compared to the force exerted by $M_*$.  In this limit, $M$ and $\mu$ free fall separately under the gravitational influence of $M_*$:
\beq
\dot{\textbf{r}}^2_1 = \frac{2 M_*}{R} \, , \qquad \dot{\textbf{r}}^2_2 = \frac{2 M_*}{|\textbf{R} - \textbf{r}|}  \, , \label{eqn: energy balance 1 2}
\eeq
and the angle between $\dot{\textbf{r}}_1$ and $\dot{\textbf{r}}_2$, denoted by $\gamma$, is given by
\beq
\cos \gamma = \frac{R^{ 2} + |\textbf{R} - \textbf{r} |^2 - r^2}{2 R |\textbf{R} - \textbf{r}|} \, .
\eeq
The dipole arising from the kinetic mixing becomes,
\beq
\begin{aligned}
 \frac{\mu}{ M \hat{\mu}} \textbf{p}_1 \cdot \textbf{p}_r  & = \mu (\dot{\textbf{r}}_1 \cdot \dot{\textbf{r}}_2 - \dot{\textbf{r}}^2_1  ) \\ & = \mu \left(   \frac{M_* (R^{ 2} + |\textbf{R} - \textbf{r} |^2 - r^2)}{(R |\textbf{R} - \textbf{r}|)^{3/2}  } - \frac{2 M_*}{R} \right)  \\[4pt]
& =  \mu \left[   \frac{M_*}{R} \left(2 + \frac{ \hat{\textbf{n}} \cdot \textbf{r}}{R} \right)  - \frac{2M_*}{R} + \mathcal{O}\left( \frac{r}{R} \right)^2 \right] \\[4pt]
& = + \left( \frac{M_* \mu }{R } \right)  \left( \frac{ \hat{\textbf{n}} \cdot \textbf{r}}{R}  \right) +  \mathcal{O}\left( \frac{r}{R} \right)^2 \, .
\end{aligned}
\eeq
As advertised, this precisely cancels the dipole contribution in (\ref{eqn: gravitational dipole}).

\newpage
\section{EFT of Extended Objects}\label{sec:eft}

In the language of effective field theory, spinning extended objects are described in terms of an effective worldline action \cite{nrgr,nrgrs,dis1,dis2, andirad,andi,prl,nrgrss,nrgrs2,nrgrso,srad1,srad2, Levi:2016ofk, eftgrg20,Goldberger:2007hy,Porto:2007pw,Foffa:2013qca,Rothstein:2014sra,Porto:2016pyg}. One of the virtues of the formalism is the inclusion of finite-size effects as a  series of higher-dimension terms beyond the `minimally coupled' point-particle worldline action, $S_{\rm pp} = M \int \d\tau$, where $M$ and $\tau$ are the mass and  proper time, respectively. One such term is the electric-type quadrupole coupling,
\beq
S_{Q} \equiv -\frac{1}{2} \int \d\tau\, Q^{ij}(\tau) E_{ij}(\tau)\,, \label{eqn: electric effective action}
\eeq
where $E_{ij}$ is the electric component of the Weyl tensor projected onto the spatial hypersurface of a free-falling basis,\footnote{The electric part of the Weyl tensor $W_{a c b d}$ is defined as $E_{ab} = W_{a c b d} u^c u^d$, with $u^a$ being the four-velocity. The tetrad is chosen such that $e^a_0 = u^a$.  Since $E_{ab}u^b=0$, only the spatial indices matter in the projection to the free-falling frame $E_{ij} = e^a_i e^b_j E_{ab}$.} and $Q^{ij}$ is the mass quadrupole moment of the object. A similar coupling appears for the magnetic component. Higher-order multipole moments can also be easily incorporated.

\vskip 4pt
The quadrupole moment can be further split into two parts:  {\it i}\hskip 1pt) a background term, which is independent of any external perturbations, and  {\it ii}\hskip 1pt) the response induced by an external field, for instance the gravitational field induced by a companion in a binary system. The effects of a permanent quadrupole moment depend on the scaling of $Q^{ij}$ with the size and spin of the body. For example, the  symmetric and trace-free spin-induced quadrupole moment of a rotating body can be parameterized as  
\beq
  \left(Q^{ij}\right)_S = -\frac{C_{ES^2}}{M} \left(S^iS^j - \frac{\delta^{ij}}{3} S^k S^k\right) ,
\eeq
where $S^i$ is the spin vector and $C_{ES^2}$ is a dimensionless Wilson coefficient that incorporates the intrinsic properties of the object. The factor of $1/M$ is chosen for convenience, so that $C_{ES^2}=1$ for Kerr BHs. This parameter is equivalent to $\kappa$, introduced in \S\ref{sec:kappa}, and measures the spin-induced quadrupolar deformability of the object, relative to the value for a rotating BH with the same mass and spin. At leading order, the quadrupole term leads to the following additional term in the worldline effective action~\cite{nrgrs,prl,nrgrss,nrgrs2}
\beq
\label{ce2}
S_{ES^2} = \frac{C_{ES^2}}{2M} \int \d\tau\,  S^{ik} S^{jk}\, E_{ij}\,,
\eeq 
where $S^{ij}$ is the spin tensor, so that $S^i = \frac{1}{2} \epsilon^{ijk} S^{jk}$. Using the power counting rules of the EFT, it is straightforward to show that this term contributes at 2PN order for binary systems with a rapidly spinning compact object (see \cite{Porto:2016pyg} for more details).

\vskip 4pt
In the presence of an external perturbation, the object's multipole moments also receive induced corrections proportional to the external field. For example, in the static limit, the response part of the quadrupole moment is\hskip 1pt\footnote{In general, the response function depends on the frequency of the external field, with $C_E$ being the real part in the low-frequency limit. The imaginary part of the response is responsible for the absorptive properties of the body. See \cite{dis1,dis2} for more details.}
\beq
\label{eq:qe}
\left(Q_{ij}\right)_{R} = -C_E\, E_{ij}\,,
\eeq 
where we have introduced the Wilson coefficient $C_E$, often called the `Love number'. 
In the worldline theory, this leads to the following higher-dimension term in the effective action,  
\beq 
S_{E^2} = \frac{C_E}{2}   \int \d\tau\, E_{ij} E^{ij} \,,
\eeq
which is quadratic in the external field. Notice that the parameter $C_E$ scales as [mass] $\times$ [size]$^4$ for general bodies, and hence as [mass]$^5$ for compact objects. It is thus conventional to introduce the dimensionless Love number, or `deformability parameter', as (see e.g.~\cite{TheLIGOScientific:2017qsa})
\beq
\Lambda \equiv \frac{C_E}{M^5}\,.
\eeq
Because of the many derivatives involved, this term enters at 5PN order for compact objects in a binary system, e.g.~\cite{nrgr}. Perhaps somewhat surprisingly, the Love numbers vanish for BHs in four-dimensional General Relativity. This unexpected `fine tuning' offers a unique opportunity to probe the nature of spacetime, and the existence of exotic objects, through precision GW data~\cite{Porto:2016zng}.

\newpage
\phantomsection
\addcontentsline{toc}{section}{References}
\bibliographystyle{utphys}
\bibliography{BHClouds-Refs}

\end{document}